\documentclass[11pt]{article}

\usepackage{array}
\usepackage{epsfig}
\usepackage{graphics}

\title{\bf A generalized Petviashvili iteration method for scalar and vector
Hamiltonian equations with arbitrary form of nonlinearity}

\author{ T.I. Lakoba\footnote{Corresponding author: lakobati@cems.uvm.edu, \ 1 (802) 656-2610} \ \ and
  J. Yang\footnote{jyang@cems.uvm.edu, \ 1 (802) 656-4314}
 \vspace{1cm} \\
  Department of Mathematics and Statistics, 16 Colchester Ave., \\
 University of Vermont, Burlington, VT 05401, USA}

\newcommand{\noi}{\noindent}
\newcommand{\D}{\Delta}
\newcommand{\be}{\begin{equation}}
\newcommand{\ee}{\end{equation}}
\newcommand{\ba}{\begin{array}}
\newcommand{\ea}{\end{array}}
\newcommand{\To}{\rightarrow}
\newcommand{\vecx}{{\bf x}}
\newcommand{\PM}{Petviashvili method}
\newcommand{\Minv}{M^{-1}}
\newcommand{\Int}{\int_{-\infty}^{\infty}}
\newcommand{\bea}{\begin{eqnarray}}
\newcommand{\eea}{\end{eqnarray}}
\newcommand{\so}{\Rightarrow}
\newcommand{\dst}{\displaystyle}

\newcommand{\tu}{\tilde{u}}
\newcommand{\tv}{\tilde{v}}
\newcommand{\bu}{\bar{u}}
\newcommand{\vecL}{{\bf L}}
\newcommand{\vecN}{{\bf N}}
\newcommand{\vece}{\vec{\bf e}}

\renewcommand{\theequation}{\thesection.\arabic{equation}}

\begin{document}
\baselineskip 18 pt

\maketitle

\vskip 2 cm

The Petviashvili's iteration method has been known as a rapidly
converging numerical algorithm for obtaining fundamental solitary
wave solutions of stationary scalar nonlinear wave equations with
power-law nonlinearity: \ $-Mu+u^p=0$, where $M$ is a positive
definite self-adjoint operator and $p={\rm const}$. In this paper,
we propose a systematic generalization of this method to both scalar
and vector Hamiltonian equations with arbitrary form of nonlinearity
and potential functions. For scalar equations, our generalized method requires only
slightly more computational effort than the original Petviashvili
method. 

\vskip 1.1 cm

\noi
{\bf Keywords}: \ Petviashvili method, Nonlinear evolution equations,
Solitary waves, Iteration methods.

\bigskip

\noi
{\bf Mathematical subject codes}: \ 35Qxx, 65B99, 65N99, 78A40, 78A99.

\newpage

\section{Introduction}

For most nonlinear wave equations arising in physical applications,
their solitary wave solutions can be obtained only numerically. The
recent interest of the research community in such applications as
Bose-Einstein condensation and light propagation in nonlinear
photonic lattices has led to a number of publications where
numerical methods for obtaining solitary waves in more than one
spatial dimension were studied. Most of these recent studies focus
on the so-called imaginary-time evolution method (ITEM), also
referred to as the normalized gradient flow method
\cite{GarciaRipollPG01, BaoD03, ShchesnovichC04, YangL06}. In this
method, one seeks a solitary wave {\em with a specified power}, or
$L_2$-norm, by numerically integrating the underlying nonlinear wave
equation with the evolution variable $t$ being replaced by $i\,t$
(hence the name `imaginary-time'). 
A key step of this
technique is the normalization of the solution's $L_2$-norm to a
given value at each iteration;
it is this step that ensures both
the convergence of the method (under known conditions
\cite{YangL06}) and the fact that the solitary wave so obtained has
a specified power\footnote{ In a modification of the ITEM, proposed
in \cite{YangL06}, one normalizes the peak amplitude (the
$L_{\infty}$-norm) of the solitary wave rather than the power. The
simulations reported in \cite{YangL06} indicate that this version of
the ITEM is faster and converges for a larger class of solutions
than the original ITEM.}. Since the power $P$ and the propagation
constant $\mu$ of a solitary wave are related 
by a beforehand unknown dependence $P=P(\mu)$, 
then the propagation constant in the ITEM 
{\em cannot} be specified and is instead computed using the available
approximation to the stationary solution at each iteration.

In some applications, it is more convenient to seek a solitary wave
with a specified propagation constant rather than with a specified
power. This is the case, for example, in nonlinear photonic
lattices, where the value of the propagation constant conveniently
parametrizes the localized solution within a spectral bandgap. One
numerical technique that can be used in this case is the Newton's
method or any of its modifications (see, e.g., \cite{ShchesnovichM04}). 
While this method is known to be very fast and also to be able to converge to both
fundamental and excited-state solitary waves, it also has drawbacks.
First, when applying the Newton's method in more than one spatial
dimension, one has to invert a matrix which is not tridiagonal. To
do so time-efficiently, one needs to use one of the alternating
direction implicit methods, which require a certain programming
effort. Second, the Newton's method often uses a finite-difference
discretization of the underlying equation, in which case the
accuracy of the obtained solution is only polynomial in $\D x$,
where $\D x$ is the typical step size of the spatial grid.
Finally, it has recently been shown that
the Newton's method may suffer erratic failures due to small
denominators \cite{Boyd05}. 
On the other hand, the ITEM mentioned in the previous
paragraph is free of these drawbacks. Namely,
the inversion of the matrix representing the differential operator
\cite{GarciaRipollPG01, YangL06} is done using the Fast Fourier
Transform, which is a built-in function in major computing software
(such as Matlab and Fortran) for one and two spatial dimensions and
can be readily extended to three dimensions. Also, since the
operator of spatial differentiation is implemented using the
spectral method, the accuracy of the ITEM is exponential in $\D x$
(provided that the solution is smooth). In addition, the ITEM does
not have the small denominator issue (although in most cases it
converges only to a dynamically stable fundamental solitary wave
\cite{YangL06}). Thus, it would be desirable to have a numerical
method that would possess the above advantages of the ITEM while
allowing the user to compute the solitary wave with a specified
value of the propagation constant rather than with the specified
power.

Such a method has long been known for a class of nonlinear wave
equations whose stationary form is
\be
-Mu+u^p=0\,,
\label{e1_01}
\ee
where $u$ is the real-valued field of the solitary wave, $M$ is a
positive definite and self-adjoint differential operator with
constant coefficients, and $p$ is a constant. For example, the
solitary wave of the nonlinear Schr\"odinger equation in $D$ spatial
dimensions,
\be
\ba{cc}
iU_t+\nabla^2 U + |U|^2U=0\,, & \quad U(|\vecx|\To \infty) \To 0\,, \vspace{0.1cm} \\
\displaystyle  \nabla^2 \equiv \frac{\partial^2}{\partial x_1^2} + \cdots
            + \frac{\partial^2}{\partial x_D^2} \,, &
\ea
\label{e1_02}
\ee
upon the substitution
\be
U(\vecx,t)=e^{i\mu t}u(\vecx)
\label{e1_03}
\ee
satisfies the equation
\be
-(\mu-\nabla^2)u + u^3=0\,,
\label{e1_04}
\ee
which has the form (\ref{e1_01}) with
\be
M=\mu-\nabla^2
\label{e1_05}
\ee
and $p=3$. Here $\mu$ is the propagation constant of the solitary
wave. We now describe the idea of the aforementioned method, which
was proposed in 1976 by V. Petviashvili \cite{Petviashvili76} and has been
referred to in the literature by his name. 
Petviashvili proposed the following iteration algorithm:
\be
u_{n+1}=\Minv u_n^p\cdot \left(
 \frac{ \langle u_n, u_n^p \rangle }{\langle u_n, Mu_n \rangle } \right)^{-\gamma}\,,
\label{e1_09}
\ee
%
where $u_n$ is the approximation of the solution at the $n$th
iteration. In scheme (\ref{e1_09}), the operators $\Minv$ and $M$ can be conveniently implemented
via the Fourier transform, e.g.:
\be
\Minv f(\vecx)\,=\, {\mathcal F}^{-1} \left[ \frac{ {\mathcal F}[f] }{ {\mathcal F}[M] }
  \right]\,,
\label{e1_07}
\ee
where
\be
\ba{l}
\displaystyle
{\mathcal F}[f] = \frac1{(\sqrt{2\pi})^D}\;
     \Int f(\vecx)\,e^{-i{\bf k}\vecx}\,d\vecx \,\equiv \hat{f}({\bf k})\,, \vspace{0.2cm} \\
\displaystyle
{\mathcal F}^{-1}[\hat{f}] = \frac1{(\sqrt{2\pi})^D}\;  
\Int \hat{f}({\bf k})\,e^{i{\bf k}\vecx}\,d{\bf k}\,,
\ea
\label{e1_08}
\ee
and $  {\mathcal F}[M] $ is the Fourier symbol of operator $M$. For example, for the operator
(\ref{e1_05}), $  {\mathcal F}[M] = \mu+{\bf k}^2 $. 
Also, the inner product in (\ref{e1_09}) and in what follows is defined in the standard way:
\be
\langle f, g \rangle = \Int f^*(\vecx)g(\vecx)\,d\vecx\,.
\label{e1_11}
\ee
(In fact, Petviashvili stated his iteration scheme using the inner product in Fourier space 
rather than its equivalent form (\ref{e1_11}).)
Note that the quotient in the parentheses of (\ref{e1_09}) equals unity when
$u_n=u$, the exact solitary wave; yet the presence of this quotient ensures the convergence of the
\PM\ when the value of the exponent $\gamma$ is taken to be in a certain range.
Namely, in \cite{Petviashvili76}, where he considered the particular case $p=2$, 
Petviashvili also formulated a mnemonic rule which yields, for any $p$, the value 
\be
\gamma=\frac{p}{p-1},
\label{e1_10}
\ee
for which the fastest convergence of the iterations (\ref{e1_09}) occurs.
The origin of this optimal value of $\gamma$ and the convergence
conditions of the \PM\ for Eq.~(\ref{e1_01}) were rigorously
established recently in Ref.~\cite{PelinovskyS04}.

The \PM\ possesses the two advantages of the ITEM which were
mentioned in the second paragraph of this Introduction. Namely, the
convenience of its implementation does not depend on the number of
spatial dimensions, and its accuracy for a smooth solution is
exponential in $\D x$. Moreover, the \PM, when it converges, is
quite fast. For example, in the case of the one-dimensional
nonlinear Schr\"odinger equation (\ref{e1_04}), if one starts with
the initial condition $u_0=e^{-x^2}$, one reaches the exact solution
with the accuracy of $10^{-10}$ in just over 30 iterations. Here and
below we define the accuracy as
\be
E_n= \left( \frac{ \langle u_n-u_{n-1}, u_n-u_{n-1} \rangle}{ \langle u_n, u_n \rangle }
  \right)^{1/2}\,.
\label{e1_12}
\ee

Recently, a number of studies reported various extensions of the
\PM\ to equations that are of a form different than (\ref{e1_01}).
In Refs. \cite{MusslimaniY04} and \cite{YangM04}, ad hoc
modifications of the Petviashvili method were proposed for the
following stationary wave equations, arising in the theory of
nonlinear photonic lattices:
\be
\nabla^2 u + V_0(\cos^2x + \cos^2y) \,u + u^3\,=\,\mu u\,,
\label{e1_13}
\ee
\be
\nabla^2 u - \frac{{\mathcal E}_0\,u}{1+V_0(\cos^2x + \cos^2y) + u^2}\,=\,\mu u\,.
\label{e1_14}
\ee
In Ref. \cite{StepanyantsT06}, another ad hoc modification of the \PM\ was proposed for the
so-called generalized Gardner equation, which has a mixed quadratic-cubic nonlinearity:
\be
\left( 1- \partial_x^2 - a \partial_y^2+\partial_x^{-2}\partial_y^2 \right) u - u^2 + bu^3=0,
\quad a>0.
\label{add1_1_01}
\ee
However, it is not straightforward to generalize the approaches of
Refs. \cite{MusslimaniY04, YangM04, StepanyantsT06} to equations
with an arbitrary form of nonlinearity. A different, systematic,
modification of the \PM\ was proposed in Ref.~\cite{AblowitzM05}.
This method, referred to in \cite{AblowitzM05} as the spectral
renormalization, can be extended to equations with
arbitrary types of nonlinearity and also to systems of coupled
equations. One can show that for a single equation with power
nonlinearity (\ref{e1_01}), the spectral renormalization method
reduces to the following scheme:
\be
u_{n+1}=\Minv u_n^p\cdot \left(
 \frac{ \langle u_n, M^{-1} u_n^p \rangle }{\langle u_n, u_n \rangle } \right)^{-\frac{p}{p-1}}\,,
\label{add2_1_01}
\ee
which is slightly different from the original \PM\ (\ref{e1_09}),
(\ref{e1_10}). (The original Petviashvili form (\ref{e1_09}) of the
spectral renormalization method can be restored if one makes a
simple modification in Eq. (6) of Ref. \cite{AblowitzM05}.)
Moreover, it can be verified that for equations that contain a
power-law nonlinear term and a potential, as, e.g.,
Eq.~(\ref{e1_13}), the spectral renormalization method with the
slight modification mentioned in parentheses above reduces to the
method of Ref.~\cite{MusslimaniY04}; see Example 3.2 in Section 3.2
for more details. However, it is not known under what conditions the
spectral renormalization method, as well as the aforementioned
methods of Refs. \cite{MusslimaniY04, YangM04, StepanyantsT06},
would converge for a general equation or a system of equations.
Also, as a minor computational issue about the spectral renormalization method,
we note that it would require some nontrivial programming effort to apply it to
equations with a non-algebraic nonlinearity, e.g., to
\be
 \nabla^2 u + \sinh u = \mu u\,.
 \label{e1_15}
\ee

In this paper, we present a generalization of the \PM\ which can be
applied to a wide class of nonlinear wave equations (including,
e.g., (\ref{e1_15})) to obtain some of their solitary wave
solutions. The idea of this generalization is based on the analysis
of the original \PM\ found in Ref. \cite{PelinovskyS04}. We also
show how our method can be applied to systems of coupled nonlinear
equations. The only restriction on the underlying physical problem
is that it be Hamiltonian. The approximate convergence conditions of
our method for a single equation are stated and discussed, and they
can be straightforwardly generalized to the case of several coupled
equations. 

The main part of this paper is organized as follows. In Section 2,
we first recast the original \PM\ into an equivalent form. All
subsequent analysis will be carried out for that equivalent
formulation of the \PM. Also in Section 2, we give a summary of the
results of Ref. \cite{PelinovskyS04} in the form that will be
suitable for a subsequent generalization. This generalization for
a single wave equation is presented in Section 3. There, we also
give examples of the applications of the new method. Next, in
Section 4, we extend this method to systems of coupled nonlinear
wave equations and present the corresponding examples. 
Thus, Sections 3 and 4 contain the two main results of this study,
which we summarize in the concluding Section 5.
The paper also contains four Appendices, whose purposes are described in
Sections 3 and 4.

The reader who is only interested in the main ideas of the generalized
Petviashvili method, but not in technical details of
its practical implementation, can skip the following material
without sacrificing the understanding: all Remarks in Section 3.1
except Remark 3.1; the entire Section 3.2; all Remarks in Section 4 except Remark 4.1;
the entire Section 4.2; and the Appendices.



\setcounter{equation}{0}
\section{Review of the analysis of the \PM\ for equations with power-law nonlinearity}

In this Section, we first reformulate the Petviashvili method into a different,
yet equivalent, form. The precise meaning of the word ``equivalent" will be stated shortly.
Then we review the results of Ref. \cite{PelinovskyS04} concerning the
convergence of the \PM\ for equations of the form (\ref{e1_01}). The way in which we present
these results is different from the way they were originally presented in \cite{PelinovskyS04}.
This reformulation of both the original \PM\ and the results of \cite{PelinovskyS04} 
will prepare the ground for our generalization of the \PM\ in Section 3.

To recast the original algorithm (\ref{e1_09}) into a
different form, let us begin by introducing a
notation. Denote the stationary equation whose solitary wave we want
to find by
\be
L_0 u=0\,.
\label{e2_01}
\ee
Thus, in the case of Eq. (\ref{e1_01}), operator
\be
L_0=-M+u^{p-1}\,.
\label{e2_02}
\ee
Here operator $M$ has the properties listed after Eq. (\ref{e1_01}),
and $u$ is the exact solitary wave. Let us rewrite the iteration
algorithm (\ref{e1_09}) in the form:
\bea
\hspace*{-1.2cm}
u_{n+1}-u_n & = & \left( u_n+\Minv [ -Mu_n+u_n^p ]\right)\, \left( 1 +
               \frac{ \langle u_n, -Mu_n+u_n^p \rangle }{ \langle u_n, Mu_n \rangle } \right)^{-\gamma}
               - u_n    \nonumber \\
 \hspace*{-1.2cm}    & = &       \left( u_n+\Minv (L_0u)_n \right)\, \left( 1 +
               \frac{ \langle u_n, (L_0u)_n \rangle }{ \langle u_n, Mu_n \rangle } \right)^{-\gamma}
               - u_n\,,
\label{e2_03}
\eea
where
$$
(L_0u)_n \equiv -Mu_n + u_n^p\,.
$$
Next, let us linearize the above equation near the exact solution $u$ by substituting into it
\be
u_n=u+\tu_n, \qquad |\tu_n| \ll |u|
\label{e2_04}
\ee
and neglecting all terms of order $O(\tu_n^2)$ and higher. Using the equation, (\ref{e2_01}),
satisfied by the solitary wave $u$, one obtains the linearized algorithm (\ref{e2_03}):
\be
\tu_{n+1}-\tu_n = \left( \Minv L\tu_n - \gamma \frac{\langle u,L\tu_n \rangle }{\langle u, Mu\rangle} u
                  \right) \,\D \tau\,,
\label{e2_05}
\ee
\be
\D \tau = 1\,,
\label{e2_06}
\ee
where $L$ is the operator of the linearized Eq. (\ref{e2_01}):
\be
L\tu_n \equiv (-M+p\,u^{p-1})\tu_n\,.
\label{e2_07}
\ee
We now interpret Eq. (\ref{e2_05}) as the explicit Euler discretization of the following
{\em continuous} linear flow:
\be
\tu_{\tau}=\Minv L\tu - \gamma \frac{\langle u,L\tu \rangle }{\langle u, Mu\rangle} u\,,
\label{e2_08}
\ee
where $\tau$ is the auxiliary (nonphysical) ``time" variable. In the
last and key step of this derivation, we ``de-linearize" the above
continuous flow:
\be
\bu_{\tau}=\Minv L_0\bu - \gamma \frac{\langle \bu,L_0\bu \rangle }{\langle \bu, M\bu\rangle} \bu\,,
\label{e2_09}
\ee
where the notation $\bu$ simply signifies that this variable is the ``current" approximation
to the exact solitary wave $u$. That is, if one linearizes Eq. (\ref{e2_09}) via a continuous analogue
of (\ref{e2_04}), one will obtain Eq. (\ref{e2_08}). Finally, we discretize Eq. (\ref{e2_09})
in time using the explicit Euler method:
\be
u_{n+1}-u_n= \left( \Minv (L_0u)_n -
  \gamma \frac{\langle u_n,(L_0u)_n \rangle }{\langle u_n, Mu_n\rangle} u_n\right)\,\D \tau\,,
\label{e2_10}
\ee
where $(L_0u)_n$ is defined after Eq. (\ref{e2_03}). Algorithm (\ref{e2_10})
with $\D \tau=1$ is equivalent to
the original Petviashvili algorithm (\ref{e1_09}) in the sense that
the linearizations of both algorithms yield the same pair of
equations (\ref{e2_05}) and (\ref{e2_06}). In the remainder of this
paper we will, therefore, refer to algorithm (\ref{e2_10}) also as
the \PM. Moreover, the {\em generalized} \PM proposed in Section 3, 
which is one of two main results of this paper, will
be based on this reformulated version of the original algorithm (\ref{e1_09}). 

Let us point out two reasons why Eq.~(\ref{e2_10}) is preferred over
Eq.~(\ref{e1_09}) for the subsequent generalization of the method.
First, the ability to select the value of the new parameter $\D\tau$ 
makes the convergence conditions of scheme (\ref{e2_10}) more relaxed
than those of the original scheme (\ref{e1_09}), as shown by Eq.~(\ref{e2_24})
below; see also a related discussion about the ITEM in \cite{YangL06}.
Second, when trying to generalize algorithm
(\ref{e1_09}), one may encounter the situation where the quotient in
the parentheses is negative and hence cannot be raised to a
non-integer power (without making $u_{n+1}$ complex-valued);
see, e.g., \cite{MusslimaniY04,YangM04}.
In contrast, algorithm
(\ref{e2_10}) is free of this difficulty. 

We now come to the second part of this Section where we will 
review those of the calculations of Ref.~\cite{PelinovskyS04} which
are essential for our own analysis given in the next Section. 
Specifically, we will exhibit the conditions under which iterations
(\ref{e2_10}) converge to the solitary wave $u$, that is, when
the error $\tu_n$ tends to zero as $n\To\infty$.
To find out when this occurs, one substitutes
the following decomposition of $\tu_n$ into (\ref{e2_05}):
\be
\tu_n(\vecx)=a_n u(\vecx) +z_n(\vecx)\,,
\label{e2_11}
\ee
where $a_n$ is a scalar (i.e., not a function of $\vecx$) and $z_n(\vecx)$ is chosen to be orthogonal to
$Mu$ at every iteration:
\be
\langle z_n, Mu\rangle =0, \quad {\rm or} \quad \langle Mz_n, u\rangle =0 \qquad
\mbox{for all $n$}.
\label{e2_12}
\ee
The second orthogonality relation follows from the first one because, by our assumption,
$M$ is a self-adjoint operator.

Before we proceed, let us first point out a relation that will be of
crucial importance both for the remainder of this section and for
Sections 3 and 4. Namely, for Eq. (\ref{e1_01}), we use Eqs. (\ref{e2_07}),
(\ref{e2_02}), and (\ref{e2_01}) to obtain:
%
\be
Lu=(p-1)Mu,
\label{e2_13}
\ee
or, equivalently,
\be
\Minv Lu=(p-1)u.
\label{e2_14}
\ee
Thus, $u$ is an eigenfunction of operator $\Minv L$, which is closely related to 
the operator on the r.h.s. of Eq. (\ref{e2_05}). 
Equation (\ref{e2_14}) is the key relation mentioned above; establishing its
counterpart for a more general Eq. (\ref{e3_01}) below will correspondingly be
one of the key steps in the generalization of the \PM\ in Section 3.
Let us note that from Eqs.
(\ref{e2_12}) and (\ref{e2_13}) there follow the orthogonality
relations
\be
\langle z_n, Lu\rangle =0, \quad {\rm or} \quad \langle Lz_n, u\rangle =0 \qquad
\mbox{for all $n$}.
\label{e2_15}
\ee
Here we have used the fact that $L$ is self-adjoint.

We now continue with the analysis of the evolution of the error $\tu_n$ with $n$. Substituting
decomposition (\ref{e2_11}) into Eq. (\ref{e2_05}) and using relation (\ref{e2_13}) and the
second of the orthogonality conditions (\ref{e2_15}), one obtains:
\be
(a_{n+1}-a_n)u+(z_{n+1}-z_n)=\Minv L z_n\,\D \tau + a_n u (p-1)(1-\gamma)\,\D \tau\,.
\label{e2_16}
\ee
Taking the inner product of this equation with $Mu$ and using the
orthogonality conditions (\ref{e2_12}) and (\ref{e2_15}), one gets
%
\be
a_{n+1}=a_n\left( 1 + (p-1)(1-\gamma)\,\D \tau\,\right)\,.
\label{e2_17}
\ee
Thus, when
\be
\gamma=1+\frac1{(p-1)\D \tau}\,,
\label{e2_18}
\ee
$a_{n+1}=0$, i.e. the component of the error $\tu_{n+1}$ ``along" the eigenfunction $u$ is
zero (in the order $O(\tu_n)$), no matter what this component was at the $n$th iteration.
Note that for $\D \tau=1$, formula (\ref{e2_18}) yields the optimal value (\ref{e1_10}) of $\gamma$
found empirically by Petviashvili.

When $a_n$ and $a_{n+1}$ are related by expression (\ref{e2_17}) (for any $\gamma$), the component
$z_n$ of the error satisfies:
\be
z_{n+1}=\left( 1+\D \tau\,\Minv L\right) z_n\,.
\label{e2_19}
\ee
Since $L$ is self-adjoint and $M$ both positive definite and
self-adjoint, eigenfunctions $\psi_j$ of $\Minv L$, satisfying
eigen-relations
\be
\Minv L \psi_j=\lambda_j \psi_j\,,
\label{e2_20}
\ee
form a complete set in the space of square-integrable functions 
and are mutually orthogonal to each other with weight $M$:
\be
\langle \psi_j, M\psi_k \rangle = \delta_{jk}\,.
\label{e2_21}
\ee
Then $z_n$ and $z_{n+1}$ can be expanded over this set:
\be
z_n=\sum_{j,\;\psi_j\neq u} Z_{j,n}\psi_j\,,
\label{e2_22}
\ee
where $Z_{j,n}$ are the expansion coefficients. The term with $\psi_j=u$ (see (\ref{e2_14}))
is excluded from the above sum because $z_n$ is orthogonal to $u$ with weight $M$ (see (\ref{e2_12})
and (\ref{e2_21})). Thus, if the eigenvalue $(p-1)$, corresponding to the eigenfunction $u$, is
{\em the only positive eigenvalue} of $\Minv L$, then $z_n$ is expandable only over the eigenfunctions with
nonpositive eigenvalues $\lambda_j$, and the expansion coefficients satisfy
\be
Z_{j,n+1}=(1+\lambda_j\D \tau)Z_{j,n}\,.
\label{e2_23}
\ee
As long as $\D \tau$ is taken sufficiently small to ensure that
\be
1+\lambda_{\min}\D \tau > -1\,,
\label{e2_24}
\ee
then $|Z_{j,n}|\To 0$ as $n\To \infty$, and hence $\lim_{n\To\infty} |z_n|=0$.
Given that $a_n=0$ (in the order $O(\tu_n)$) at every iteration when $\gamma$ is chosen
according to (\ref{e2_18}), decomposition (\ref{e2_11})  implies that $|\tu_n|\To 0$ as
$n\To \infty$. That is, the \PM, under the above conditions, converges to the solitary wave $u$.



\setcounter{equation}{0}
\section{The generalization of the \PM\ for a single nonlinear wave equation with
a general form of nonlinearity}

This section contains the first main result of this study. Namely, 
we will show how the Petviashvili method can be generalized for an equation of the form
\be
L_0 u \equiv -Mu+F(\vecx,u) =0, \qquad u(|\vecx|\To \infty)\,\To 0,
\label{e3_01}
\ee
where $F(\vecx,u)$ is any real-valued function. 
In Section 3.1, we will derive and discuss the algorithm of this
method and in Section 3.2 will illustrate it with examples.

\subsection{Derivation of the generalized Petviashvili method}

Recall that one of the key results of Section 2 was Eq.
(\ref{e2_13}). It was that relation on which the usefulness of decomposition
(\ref{e2_11}) was based; see the derivations of Eqs. (\ref{e2_16})
and (\ref{e2_17}). Therefore, we will seek to obtain a counterpart
of (\ref{e2_13}) for Eq. (\ref{e3_01}). To make the main problem of
obtaining such a counterpart clearer, consider a particular case of
that equation that arises, e.g., in the theories of Bose-Einstein
condensation and light propagation in nonlinear photonic lattices:
\be
L_0 u \equiv -Mu + V(\vecx)u+u^3 =0, \qquad u(|\vecx|\To \infty)\,\To 0.
\label{e3_02}
\ee
In the aforementioned
physical applications, $M$ is given by Eq. (\ref{e1_05}), and $V(\vecx)$ is some potential.
The linearized operator $L$ in this case is
\be
L=-M+V(\vecx)+3u^2,
\label{e3_03}
\ee
and hence
\be
Lu \,=\, -Mu+V(\vecx)u+3u^3\,=\, 2u^3 \,=\, 2(M-V(\vecx))u \,\neq\,{\rm const}\cdot Mu\,.
\label{e3_04}
\ee
Thus, an {\em exact} counterpart of Eq. (\ref{e2_13}) for a general
stationary wave equation (\ref{e3_01}) {\em cannot} be obtained.

As a solution to the above problem, we propose to seek such a positive definite and self-adjoint
operator $N$ that the counterpart of (\ref{e2_13}) would hold {\em approximately}:
\be
Lu \approx \alpha Nu\,.
\label{e3_05}
\ee
Here both $N$ and the constant $\alpha$ remain to be determined.
Given such $N$ and $\alpha$, we then construct the following
counterpart of algorithm (\ref{e2_10}):
\be
u_{n+1}-u_n = \left( N^{-1} (L_0u)_n - \gamma\,
 \frac{ \langle u_n, (L_0 u)_n \rangle }{ \langle u_n, N u_n \rangle }\,u_n \right)\D \tau\,,
\label{e3_06}
\ee
where
\be
\gamma=1+\frac1{\alpha \D \tau}\,.
\label{e3_07}
\ee

The algorithm given by the iteration scheme (\ref{e3_06}), (\ref{e3_07}) is the main result 
of this section. We will refer to it as the generalized Petviashvili method.
All the steps of the analysis in
the second part of Section 2 can now be repeated, leading to the
following approximate (see below) convergence condition for this new method:

{\em If operator $N^{-1}L$ has only one positive eigenvalue (which approximately equals $\alpha$)
and if the step size $\D \tau$ satisfies inequality (\ref{e2_24}), where now $\lambda_{\min}$ is the
most negative eigenvalue of $N^{-1}L$, then the generalized \PM\ converges to the exact solitary wave
\ $u(\vecx)$. }

Two comments are in order here. First, the component of the error $\tu_n$ ``aligned
along" the eigenfunction of $N^{-1}L$ corresponding to the (only)
positive eigenvalue will be annihilated in the generalized \PM\ not completely, 
as in the original method (\ref{e1_09}), (\ref{e1_10}), but approximately. This
is due to the fact that $u$ and $\alpha$ are no longer the exact
eigenfunction and eigenvalue of $N^{-1}L$, and hence taking $\gamma$
according to Eq. (\ref{e3_07}) does not make $a_n$ in (\ref{e2_11})
exactly zero at each iteration. That, however, is not really
required for convergence: It is sufficient that
$|a_{n+1}|<|a_n|$ for all $n$, which is a much more relaxed
condition than $a_{n+1}=0$; see also Remark 3.3 below.
Second, the reason why the convergence condition stated above in italic is approximate rather than exact,
as the similar condition for the original \PM\ stated after Eq.~(\ref{e2_24}), is the following. 
Since $z_n$ in (\ref{e2_11}) is no longer exactly orthogonal to $Nu$, then the corresponding counterpart of 
Eq.~(\ref{e2_19}) for method (\ref{e3_06}), (\ref{e3_07}) will hold only approximately. Therefore, in
principle it is conceivable that if the exact eigenvalue $\lambda_2$ of $N^{-1}L$ is close to zero and
negative, the corresponding eigenvalue of the linearized operator on the r.h.s. of (\ref{e3_06}) will
be slightly positive (or vice versa). However, such cases are expected to be rare in applications of
this method. In fact, we did not encounter them in any of the equations to which we applied
algorithm (\ref{e3_06}), (\ref{e3_07}).

We will now show how the operator $N$ and constant $\alpha$ in (\ref{e3_05}) can be determined in
an efficient way. It should be noted that we cannot give the most general recipe in this regard,
simply because there are infinitely many possibilities here, as it will become clear as we proceed.
Instead, we will consider in detail only one typical case that arises in many applications and 
will show how $N$ can be found for it. At the end of this subsection we will also briefly comment
on another example of finding $N$. 

Suppose that $M$ in Eq. (\ref{e3_01}) is given by (\ref{e1_05}). The simplest
{\em ansatz} for $N$ is then
\be
N=c-\nabla^2,
\label{e3_08}
\ee
where $c$ is to be determined from the condition that ``vector" $Nu_n$ be ``aligned along"
``vector" $Lu_n$ as closely as possible. Therefore, we require that
\be
\frac{ \langle Nu_n, Lu_n \rangle ^2}{\langle Nu_n, Nu_n \rangle \langle Lu_n, Lu_n \rangle }
 \,=\, \max\,.
\label{e3_09}
\ee
Differentiating the l.h.s. of the above condition with respect to $c$ and setting the result to zero,
one obtains
\be
\frac{ \langle N_c u_n, Lu_n \rangle }{\langle N_c u_n, Nu_n \rangle }  \,=\,
\frac{ \langle N u_n, Lu_n \rangle }{\langle N u_n, Nu_n \rangle }  \,,
\label{e3_10}
\ee
where $N_c\equiv \partial N/\partial c =1$. The substitution of expression (\ref{e3_08}) into
(\ref{e3_10}) yields the value for $c$ at the $n$th iteration:
\be
c_n\,=\,
\frac{ \langle u_n,Lu_n \rangle \langle \nabla^2 u_n, \nabla^2 u_n \rangle -
       \langle \nabla^2 u_n,Lu_n \rangle \langle u_n, \nabla^2 u_n \rangle }
     { \langle u_n,Lu_n \rangle \langle u_n, \nabla^2 u_n \rangle -
       \langle \nabla^2 u_n,Lu_n \rangle \langle u_n, u_n \rangle }\,.
\label{e3_11}
\ee
It is straightforward to verify that for equations with
power-law nonlinearity (\ref{e1_01}) with $M$ of the form (\ref{e1_05}),  Eq.~(\ref{e3_11})
yields $c=\mu$ and hence $N=M$.

Now that $N$ has been determined from (\ref{e3_08}) and (\ref{e3_11}), 
the approximate eigenvalue $\alpha$ in (\ref{e3_05}) can be found from
\be
\alpha_n = \frac{ \langle u_n, Lu_n \rangle }{ \langle u_n, Nu_n \rangle } \,.
\label{e3_12}
\ee
Thus, Eqs. (\ref{e3_11}), (\ref{e3_12}), (\ref{e3_06}), and (\ref{e3_07}) provide all the
necessary information for the implementation of the generalized \PM. 

Before commenting on another example of finding $N$, we will make several remarks
regarding implementation of Eqs. (\ref{e3_11}) and (\ref{e3_12}) in a code. As noted
at the end of the Introduction, the reader who is not interested in such technical details
may read only Remark 3.1 and then proceed directly to Section 4.

\smallskip

{\bf Remark 3.1} \ There is no apriori guarantee that the constant $c$ obtained from (\ref{e3_11})
will be positive, as is required in order to make operator $N$ positive definite. However, in all of
the examples considered below we monitored $c_n$ and observed it being positive as long as we started with
a ``reasonable" initial condition $u_0$.

\smallskip

{\bf Remark 3.2} \ This concerns the calculation of quantity $Lu_n$ in Eq. (\ref{e3_11}).
Note that for any number $\kappa$,
\bea
(Lu)_n+ \kappa(L_0u)_n & \equiv &
 \left( -Mu_n + F_u(\vecx,u_n)u_n \right) + \kappa \left( -Mu_n + F(\vecx,u_n) \right)
 \nonumber \\
 & = & Lu + (Lu)_u\tu_n + \kappa\, L\tu_n + O(\tu_n^2)   \nonumber \\
 & = & Lu + O(\tu_n) \,,
\label{e3_13}
\eea
i.e., in the leading order this expression is independent of
$\kappa$. In (\ref{e3_13}), $F_u \equiv \partial F/\partial u$,
$(Lu)_u \equiv \partial (Lu)/\partial u$, and we have used the fact
that $L_0u=0$. However, in practice, the initial condition $u_0$ may
not be ``sufficiently close" to the exact solution $u$. This will
make the $O(\tu_n)$-correction comparable in size with the first
term on the r.h.s. of (\ref{e3_13}), which will affect the values of
$c_n$, $\alpha_n$, and $\gamma_n$ in (\ref{e3_11}), (\ref{e3_12}),
and (\ref{e3_07}). This, in turn, may prevent the algorithm from
converging. In our simulations, we found that this can indeed occur.
Then we found, empirically, that calculating $Lu_n$ in (\ref{e3_11})
by using (\ref{e3_13}) with $\kappa=-1$, i.e.,
\be
Lu_n \equiv (Lu)_n - (L_0u)_n  = F_u(\vecx,u_n)u_n - F(\vecx,u_n)\,,
\label{e3_14}
\ee
greatly increases the range of the initial conditions $u_0$ for which the above algorithm converges.
For example, in the case of Eq. (\ref{e3_02}), \ $Lu_n=2u_n^3$.

\smallskip

{\bf Remark 3.3} \ The approximate eigenvalue $\alpha$ can be calculated by any formula that is
equivalent to (\ref{e3_12}) had (\ref{e3_05}) held exactly rather than approximately. For example,
an alternative to (\ref{e3_12}) may be taken as
\be
\alpha_n = \frac{ \langle Nu_n, Lu_n \rangle }{ \langle Nu_n, Nu_n \rangle }\,.
\label{e3_15}
\ee
However, in all the examples considered below, we found that the
convergence rate was the same no matter whether (\ref{e3_12}) or
(\ref{e3_15}) had been used. This is so because the value of
$\alpha$ affects only the value of $\gamma$, which, in its turn,
determines how far the ratio $(a_{n+1}/a_n)$ is from zero. But if
this ratio is, say, $0.2$ instead of $0.05$ (or vice versa) due to a
slight variation in $\gamma$, this does not affect the convergence
rate, since the latter is determined, in most if not all cases, by
the much slower decay of $\| z_n\| \equiv \sqrt{ \langle z_n, z_n
\rangle }$.

\smallskip

{\bf Remark 3.4} \ For some equations, $\alpha$ can be quite small (say, on the order of $0.01$ or less).
We encountered such cases among the examples considered in Section 4. A small $\alpha$ yields a large
value of $\gamma$; see (\ref{e3_07}). We found it to be beneficial to artificailly limit such large
$\gamma$'s by using, e.g.,
\be
\gamma= \frac{ \gamma_{\rm aux} }{\sqrt{1+( \gamma_{\rm aux} / \gamma_{\max} )^2 } } \,,
\label{e3_17}
\ee
where $\gamma_{\rm aux}$ is defined by the r.h.s. of (\ref{e3_07}) and $\gamma_{\max}$ is some large
number specific to the problem at hand. The reason why such a limiting may be needed is as follows.
Since $u$ is not an exact eigenfunction of $N^{-1}L$, is can be represented as
$$
u=U_1\psi_1 + \sum_{j\ge 2} U_j\psi_j\,,
$$
where $\psi_j$ are the true eigenfunctions of $N^{-1}L$, and the expansion coefficients $U_j$ are
such that $|U_j|\ll |U_1|$ for $j\ge 2$. That is, $u$ contains ``small pieces" of eigenfunctions
other than $\psi_1$ (the latter is the eigenfunction that $u$ approximates). While the value of $\gamma$
as given by (\ref{e3_07}) is chosen so as to annihilate the $\psi_1$-component of the error $\tu_n$,
it is not intended to annihilate any of the other $\psi_j$-components. On the contrary, if
$\gamma$ is ``too large", this may {\em amplify} some of those components, which would then result in the
divergence of the iterations. Obviously, this could not have occurred in the case of Eq. (\ref{e1_01}),
since there $u=\psi_1$ and thus all $U_j=0$ for $j\ge 2$.

\smallskip

{\bf Remark 3.5} \ Finally, we note that the computation of $c_n$ at
every iteration slows down the execution of the code because such a
computation requires evaluation of the inner products $\langle u_n,
Lu_n \rangle$ and $\langle \nabla^2 u_n, Lu_n \rangle$, which are
not used in the iteration equation (\ref{e3_06}) itself. However, by
the same argument as in Remark 3.3, it is sufficient to compute
$c_n$, $\alpha_n$, and $\gamma_n$ only until the solution reaches
some low accuracy (defined by Eq. (\ref{e1_12})), say, $10^{-3}$,
and then carry on the rest of the iterations using the values
of $c_n$ and $\gamma_n$ computed up to that moment.

\bigskip

As a case where a more involved ansatz for $N$ than (\ref{e3_08})
may appear to be more appropriate,
 consider Eq. (\ref{e3_02}) in two spatial dimensions where $M$ is given by
(\ref{e1_05}), i.e., is isotropic in the spatial dimensions, but the
potential $V(\vecx)$ is essentially anisotropic. In this case, one
may expect that an ansatz more general than (\ref{e3_08}), namely,
\be
N=c-(b\,\partial_x^2 + \partial_y^2)\,,
\label{e3_18}
\ee
would allow the approximate equation (\ref{e3_05}) to hold with a
better accuracy, which, in turn, may result in faster convergence of
the iterations. However, this turns out not to be so in general.
Specifically, we used ansatz (\ref{e3_18}) for finding solitary waves
of equation
\be
\nabla^2 u + V_0 \left( {\rm sech}^2(3x) - 1\right) u + u^3 = \mu u\,
\label{e3_27}
\ee
(where the potential depends on $x$ but not on $y$)
and observed that not only does using this more involved ansatz require more
coding effort than when using the simpler ansatz (\ref{e3_08}),
 but it also leads to slower convergence of the iterations.
To conserve the printed space and the reader's time, we do not show an 
analysis of this case, since it apparently has little practical usefulness.

We now present three examples that demonstrate the validity of our algorithm
(\ref{e3_06})--(\ref{e3_08}), (\ref{e3_11}), (\ref{e3_12}).


\subsection{Examples of application of the generalized \PM\ to a single nonlinear wave equation}

{\bf Example 3.1} \ We apply our method to equation
\be
\nabla^2 u + V_0(\cos^2x+\cos^2y)u+u^3=\mu u
\label{e3_21}
\ee
with $V_0=3$ nd $\mu=3.7$. The initial condition for the iterations is
\be
u_0=A\,e^{-(x^2+y^2)/W^2}
\label{e3_22}
\ee
with $A=1$ and $W=1$. We take the free parameter $\D \tau=1$ and
compute $c$ and $\gamma$ until the solution reaches the accuracy of
$10^{-3}$ (see Remark 3.5). The latest computed values are $c=1.20$ and
$\gamma=3.71$. Then the iterations are continued until the solution
reaches the accuracy of $10^{-10}$. The final solution is shown in Fig.~\ref{fig1},
and a short Matlab code that can be used to obtain it is
given in Appendix 1. 

The total number of iterations taken by the generalized \PM\ 
is about 180. (Here and below we quote the
number of iterations rounded to the nearest ten. The reason is that
this number may slightly depend on the size of the computational
domain and possibly other technical factors.) 
For comparison, the
optimally accelerated ITEM (with the corresponding power of the
solitary wave being $P=3.0$) reaches the same solution in about 300
iterations; see Example 9.1 in \cite{YangL06}. The modification of
the ITEM where one seeks a solitary wave with a specified peak
amplitude \cite{YangL06} rather than the power converges to the same
solution in 130 iterations. The ad hoc modification of the original
\PM, proposed for Eq. (\ref{e3_21}) in \cite{MusslimaniY04}, takes
about 420 iterations to converge to the same accuracy
\cite{YangL06}. It should be noted that the value $\D \tau=1$, which
we used in this Example, does {\em not} lead to the fastest
convergence of algorithm (\ref{e3_06})--(\ref{e3_08}). For instance,
we found that for $\D \tau=1.3$, the convergence rate of our method
is nearly the fastest, and the method converges in about 140
iterations. The dependence of the convergence rate on the step size $\D t$ is
discussed in the companion paper \cite{paper2}.

\smallskip

{\bf Example 3.2} \ In this Example we present a case where two
modifications of the original \PM\ proposed in Refs. \cite{MusslimaniY04} and \cite{AblowitzM05} diverge,
but the generalized \PM, proposed in this Section,
converges. We seek an anti-symmetric solution of the following equation with a double-well potential:
\be
u_{xx}+ V(x)u - u^3 = \mu u , \qquad V(x)=6\left( {\rm sech}^2(x-1) + {\rm sech}^2(x+1) \right) 
\label{add3_01}
\ee
for $\mu=1.43$. 
(The solution with this value of the propagation constant has the power $P\equiv \Int u^2\,dx=10$
and was originally found in \cite{YangL06} by the ITEM.)
As the initial condition, we take 
\be
u_0=2x\,e^{-x^2} + \epsilon \, e^{-x^2},
\label{add3_02}
\ee
with $\epsilon$ being either zero or $0.001$. 
As in Example 3.1, we compute $c$ and
$\gamma$ only as long as the error exceeds $10^{-3}$; then
we use these latest computed values for the rest of the iterations. 
We first set $\epsilon=0$ in (\ref{add3_02}) and empirically find that
$\D \tau=1.6$ results in the fastest convergence (in about 40 iterations) 
of the generalized \PM\
(\ref{e3_06})--(\ref{e3_08}); the iteratively computed parameters of the
algorithm are in this case: $c=5.04$ and $\gamma=0.21$. 
The corresponding solution 
is shown in Fig.~\ref{fig2}.  Next, when we introduce a small symmetric
component into the initial condition (\ref{add3_02}) by setting
$\epsilon=0.001$, the iterations still converge to that solution, although at
a lower rate (in about 170 iterations). 

We now apply to Eq. (\ref{add3_01}) the modifications of the original \PM\ proposed in Refs.
\cite{MusslimaniY04} and \cite{AblowitzM05}. The former of these methods has the form:
\be
u_{n+1}= M^{-1} \left( C_n^{\gamma_{\rm lin}} V(x)u_n - C_n^{\gamma_{\rm nl}} u_n^3 \right),
\label{add3_03}
\ee
where $M=\mu-\partial_x^2$
and the factor $C_n$ is chosen so that it equals one when $u_n$ is an exact solution of (\ref{add3_01}):
\be
C_n= \frac{ \langle u_n, \, (-M+V(x))u_n \rangle }{ \langle u_n, \, u_n^3 \rangle } .
\label{add3_04}
\ee
The constants $\gamma_{\rm lin}$ and $\gamma_{\rm nl}$ in (\ref{add3_03}) are to be chosen empirically;
in \cite{MusslimaniY04}, the choice $\gamma_{\rm lin}=0.5$ and $\gamma_{\rm nl}=1.5$ was suggested.
The method of Ref.~\cite{AblowitzM05} for Eq.~(\ref{add3_01}) can be shown to reduce to the same form
(\ref{add3_03}), where now
\be
C_n= \frac{ \langle u_n, \, (-1+M^{-1}V(x))u_n \rangle }{ \langle u_n, \, M^{-1}u_n^3 \rangle }
\label{add3_05}
\ee
and $\gamma_{\rm lin}=0.5$ and $\gamma_{\rm nl}=1.5$. (Unlike in the
ad hoc method of Ref.~\cite{MusslimaniY04} where these values of
$\gamma_{\rm lin}$ and $\gamma_{\rm nl}$ were ``guessed", in the
method of Ref.~\cite{AblowitzM05} these values can be derived from
Eqs.~(5) and (6) of that paper.)
For the solution of (\ref{add3_01}) with $\mu=1.43$, both these
methods converged in about 20 iterations when started at the initial
condition (\ref{add3_02}) with $\epsilon=0$. However, they both diverged for
$\epsilon=0.001$, in contrast to our generalized \PM, which converged for either
value of $\epsilon$.

\smallskip

{\bf Example 3.3} \ In this Example, we show that the calculations
of Sections 2 and 3 for the optimal value of $\gamma$ can be carried
out even when the nonlinearity of the equation is nonlocal. Consider
a stationary wave equation
\be
\left( 1 - \frac12 \nabla^2 \right) u \, = \, u\,
\nabla^{-2} \left( \frac{\partial^2}{\partial x^2} u^2 \right).
\label{e3_23}
\ee
This equation, which has a nonlocal operator $\nabla^{-2}$ on the right hand side, can be rewritten
as a system of two local equations:
\be
\ba{l}
\dst
\left( 1 - \frac12 \nabla^2 \right) u \, = \, u\,v, \vspace{0.1cm} \\
\dst
\nabla^2 v = \frac{\partial^2}{\partial x^2} u^2 \,.
\ea
\label{e3_24}
\ee
This system arises as the small-field approximation in the theory of light propagation in
photorefractive media; see, e.g., \cite{ZozulyaA98}. Let us note that Eqs. (\ref{e3_24}) cannot
be handled by the method described in Section 4 below because the corresponding linearized operator
is not self-adjoint. However, the original nonlocal Eq. (\ref{e3_23}) can be handled by the method
of Sections 2 and 3. To that end, let us linearize this equation near the exact solution $u$:
\be
L\tu \equiv - \left( 1 - \frac12 \nabla^2 \right) \tu +
\tu\, \nabla^{-2} \left( \frac{\partial^2}{\partial x^2} u^2 \right) +
u \nabla^{-2} \left( \frac{\partial^2}{\partial x^2} 2u\tu \right).
\label{e3_25}
\ee
Then, using (\ref{e3_23}),
\be
Lu \,=\, 2u \nabla^{-2} \left( \frac{\partial^2}{\partial x^2} u^2 \right) \,=\,
2 \left( 1 - \frac12 \nabla^2 \right) u\,.
\label{e3_26}
\ee
This is Eq. (\ref{e2_13}) with $p=3$ and $M=\left( 1 - \frac12
\nabla^2 \right)$; hence, from (\ref{e1_10}), $\gamma_{\rm
opt}=3/2$. It is this value of $\gamma$ which the authors of
\cite{ZozulyaA98} used (without justification) in the original
Petviashvili algorithm applied to system (\ref{e3_24}).



\setcounter{equation}{0}
\section{Generalization of the \PM\ for coupled nonlinear wave equations}

Here we will first show how the generalized \PM\ of Section 3 can be
extended to obtain solitary waves in Hamiltonian systems of
coupled nonlinear equations. Then we will present the
corresponding examples. To make the essential details of our
technique clearer, we will focus on the case of two coupled
equations, while commenting on the extension to three
and more equations in Appendix 3.

\subsection{Derivation of the generalized \PM\ for coupled equations}

Consider the following system of equations for the real-valued
components $u$ and $v$ of the solitary wave: 
\be
 - \left(  \ba{cc} M_{11} & 0 \\ 0 & M_{22} \ea \right)
 \left( \ba{c} u \\ v \ea \right) + \left( \ba{c} F_1(\vecx,u,v) \\ F_2(\vecx,u,v) \ea \right)
 \; \equiv \; \vecL_0 \left( \ba{c} u \\ v \ea \right) \; = \; \left( \ba{c} 0 \\ 0 \ea \right)\,,
\qquad  \lim_{|\vecx|\To \infty} \left( \ba{c} u \\v \ea \right) = \left( \ba{c} 0 \\ 0 \ea \right),
\label{e4_01}
\ee
where $M_{11}$ and $M_{22}$ are self-adjoint positive definite operators. (Whenever symmetric (see below)
off-diagonal terms $M_{12}v$ and $M_{12}u$ are present in the first and second equations, 
they can always be removed by a linear transformation.) 
The restrictions on functions $F_{1,2}$ become clear when
one considers the linearized operator, $\vecL$, of Eq. (\ref{e4_01}):
\be
\vecL \left( \ba{c} \tu \\ \tv \ea \right) =
\left( \ba{cc} -M_{11}+F_{1,u} & F_{1,v} \\ F_{2,u} & -M_{22}+F_{2,v} \ea \right)
\left( \ba{c} \tu \\ \tv \ea \right) \,,
\label{e4_02}
\ee
where $F_{1,u}\equiv \partial F_1 /\partial u$, etc. Recall that the linearized operator $L$ played
the key role in the analysis of Section 2; in particular, it was crucial for the derivation of Eqs.
(\ref{e2_15})--(\ref{e2_17}) and the discusion following Eq. (\ref{e2_19}) that $L$ was self-adjoint.
Similarly, to carry out that analysis for the coupled Eqs. (\ref{e4_01}), we require that $\vecL$
in (\ref{e4_02}) be self-adjoint. This yields the condition
\be
F_{1,v}=F_{2,u}\,.
\label{e4_03}
\ee
Thus, our method will be applicable to systems of form (\ref{e4_01}) where $F_1$ and $F_2$ satisfy
condition (\ref{e4_03}). Note that these functions may contain nonlocal operators as in Example 3.3
above.

Our plan now is as follows. We will first present a generalization of the linearized continuous
flow (\ref{e2_08}), then will comment on it, and, finally, will state the vector counterpart of the
``delinearized" algorithm (\ref{e3_06}). The extension of (\ref{e2_08}) to the vector case is:
\be
 \left( \ba{c} \tu \\ \tv \ea \right)_{\tau} = \vecN^{-1} \vecL \left( \ba{c} \tu \\ \tv \ea \right)
 - \sum_{k=1}^2 \gamma_k
\frac{ \left\langle \vece_k, \, \vecL \left( \ba{c} \tu \\ \tv \ea \right) \right\rangle }
{\langle \vece_k,\, \vecN \vece_k \rangle }\,\vece_k\,,
\label{e4_04}
\ee
\be
\gamma_k=1+\frac1{\alpha_k \D \tau}\,, \qquad
\alpha_k = \frac{\langle \vece_k,\, \vecL \vece_k \rangle }{\langle \vece_k,\, \vecN \vece_k \rangle }\,,
\qquad k=1,2,
\label{e4_05}
\ee
where $\vecN$ is a self-adjoint, positive definite matrix operator, whose form will be discussed
shortly, and $\vece_k$ and $\alpha_k$ are the approximate eigenvectors and eigenvalues
of $\vecN^{-1}\vecL$:
\be
\vecL\,\vece_k\, \approx \, \alpha_k \vecN\,\vece_k\,.
\label{e4_06}
\ee
The analysis of convergence of Eq. (\ref{e4_04}) proceeds along the lines of the corresponding analysis
in Section 2 with one minor modification:
In the derivation of Eq. (\ref{e4_05}) for $\gamma_k$, one needs to use the (approximate)
orthogonality of $\vece_1$ and $\vece_2$:
\be
\langle \vece_1, \, \vecN \vece_2 \rangle =0\,.
\label{e4_09}
%
%
\ee
Condition (\ref{e4_09}) follows from (\ref{e4_06}) and the fact that both $\vecL$ and $\vecN$
are self-adjoint.

Now we discuss the computationally efficient choice of operator $\vecN$ and vectors $\vece_k$.
It is this choice that makes the generalization of the method of Section 3 to the case of
coupled equations nontrivial; hence it constitutes an important technical result of this Section.
For the simplicity of presentation, we assume that both $M_{11}$ and $M_{22}$ in (\ref{e4_02})
have form (\ref{e1_05}), with possibly different $\mu$'s. (The extension to a more general form
of these operators is straightforward and one instance of it is given in Example 4.3 below.)
Then, the form of $\vecN$ that we advocate, and which we used in all of the examples presented
in Section 4.2, is
%
%
%
\be
\vecN = \left( \ba{cc} N_1 & 0 \\ 0 & N_2 \ea \right) , \qquad
N_k=c_{k}-b_{k}\nabla^2, \quad k=1,2.
\label{e4_11}
\ee
One may wonder if the more general form that includes (symmetric) off-diagonal terms 
$c_{12}-b_{12}\nabla^2$ would result in a more efficient 
method. The answer, based on our experimentation with both this more general form and the
simpler form (\ref{e4_11}), is negative. First, the coding of the part of the
computer program that would calculate all of the coefficients $c_k$, $b_k$, and $c_{12}$, $b_{12}$
for the more general form of $N$ is considerably more tedious than the corresponding coding for
the simpler form (\ref{e4_11}). This part of the program would be difficult to debug had a mistake 
in it occurred. Moreover, the simplicity of the original Petviashvili method, which is one of its 
main advantages over the Newton's method, would be compromised by this coding issue. Second,
in our simulations we also found that, in some cases, unless the initial condition $(u_0,v_0)$ is 
``very" close to the exact solitary wave $(u,v)$, then the $\vecN$ calculated as a full matrix 
may turn out not to be positive definite, which would
result in the divergence of the iterations. On the other hand, we verified that the simpler,
diagonal form (\ref{e4_11}) does not have either of the above drawbacks.

Let us now show how $c_{1,2}$ and $b_{1,2}$ in (\ref{e4_11}) can be computed while assuming a
general form of the eigenvector $\vece_1$, and then will argue that one can and should take
$\vece_1=(u,v)^T$. Let
\be
\vecL \equiv \left( \ba{cc} L_{11} & L_{12} \\ L_{12} & L_{22} \ea \right),
\qquad
\vece_1 \equiv \left( \ba{c} e_{11} \\ e_{21} \ea \right),
\label{e4_12}
\ee
where each of $L_{ij}$ is a self-adjoint operator. As in Section 3, we require that
\be
\frac{ \langle \vecN\vece_1 , \, \vecL\vece_1 \rangle^2 }
{  \langle \vecN\vece_1 , \, \vecN\vece_1 \rangle  \langle \vecL\vece_1 , \, \vecL\vece_1 \rangle }
\, =\, \max
\label{e4_13}
\ee
and then find the equations for $c_{1,2}$ and $b_{1,2}$ by setting the derivatives of the l.h.s.
with respect to these parameters to zero. Thus, similarly to (\ref{e3_10}), one obtains a system
of four equations:
\be
\frac{\langle \vecN_r\vece_1 , \, \vecL\vece_1 \rangle}{\langle \vecN_r\vece_1 , \, \vecN\vece_1 \rangle}
 \,=\,
\frac{\langle \vecN\vece_1 , \, \vecL\vece_1 \rangle}{\langle \vecN\vece_1 , \, \vecN\vece_1 \rangle},
\label{e4_14}
\ee
$$
\vecN_r \equiv \frac{\partial \vecN }{\partial r}, \qquad r=\{ c_1,c_2,b_1,b_2\}\,.
$$
Since the r.h.s. of all these equations is the same, one can obtain a system of three
equations for the four unknown parameters $r$ by setting the correponding l.h.s.'s equal to each
other. It is easy to see, by inspection, that this system is {\em linear} and homogeneous, and hence
it produces a solution for $\{c_1,c_2,b_1,b_2\}$ that is unique up to multiplication by an arbitrary
constant. (Such an arbitrariness is expected because operator $\vecN$ is defined by (\ref{e4_06}) only
up to an arbitrary factor.) Next, any one of Eqs. (\ref{e4_14}) can be taken as the remaining fourth
equation for $\{c_1,c_2,b_1,b_2\}$. We verified that such an equation is satisfied identically for
the solution $\{c_1,c_2,b_1,b_2\}$ determined from the aforementioned linear homogeneous system of three
equations.

The solution of that system can most easily be found as follows. Equating the l.h.s.'s of Eqs.
(\ref{e4_14}) with $r=c_k$ to those with the corresponding $r=b_k$ yields:
\be
\kappa_k \equiv \frac{c_k}{b_k} =
\frac{ \left\langle \left( \,\langle \nabla^2 e_{k1},\, \sum_{j=1}^2 L_{kj} e_{j1} \rangle\, e_{k1} -
 \langle e_{k1},\, \sum_{j=1}^2 L_{kj} e_{j1} \rangle \,\nabla^2 e_{k1} \right), \, \nabla^2 e_{k1}
 \right\rangle }
{ \left\langle \left( \,\langle \nabla^2 e_{k1},\, \sum_{j=1}^2 L_{kj} e_{j1} \rangle\, e_{k1} -
 \langle e_{k1},\, \sum_{j=1}^2 L_{kj} e_{j1} \rangle \,\nabla^2 e_{k1} \right), \, e_{k1}
 \right\rangle } \,, \qquad k=1,2.
\label{e4_15}
\ee
Then, equating the l.h.s.'s of (\ref{e4_14}) with $r=c_1$ and $r=c_2$ yields
\be
\frac{b_2}{b_1} =
\frac{ \langle e_{11}, \, (\kappa_1-\nabla^2)e_{11} \rangle  \;
       \langle e_{21}, \, \sum_{j=1}^2 L_{2j}e_{j1} \rangle }
{ \langle e_{21}, \, (\kappa_2-\nabla^2)e_{21} \rangle  \;
       \langle e_{11}, \, \sum_{j=1}^2 L_{1j}e_{j1} \rangle } \,.
\label{e4_16}
\ee
The pseudocode for the time-efficient computation (i.e., a computation that avoids repeated evaluation
of the same quantities) is presented in Appendix 2. As in Remark 3.5, we note that $c_{1,2},\;b_{1,2}$
and the corresponding values of $\alpha_{1,2}$ and $\gamma_{1,2}$ need only be computed up to the moment
when the iterations approach the exact solution with some relatively low accuracy (say, $10^{-3}$).
The remaining iterations, up to a higher accuracy, can be carried out with those latest computed values
of these parameters.

{\bf Remark 4.1} \ Let us reiterate that the above algorithm of finding the coefficients of operator
$\vecN$, which can be straightforwardly generalized to any number of coupled equations (see
Appendix 3), is one of
the main results of this Section. The key part here is that a unique set of these coefficients
can {\em always} (except, maybe, in some pathological cases which we never encountered)
be found by solving a {\em linear} system of equations.

We now discuss the choice of the eigenvectors $\vece_1$ and $\vece_2$. First, we note that since
these eigenvectors enter Eq. (\ref{e4_04}) on equal footing, it might seem that it would be ``more correct"
to replace the l.h.s. of (\ref{e4_13}) by
\be
\sum_{k=1}^2
\frac{ \langle \vecN\vece_k , \, \vecL\vece_k \rangle^2 }
{  \langle \vecN\vece_k , \, \vecN\vece_k \rangle  \langle \vecL\vece_k , \, \vecL\vece_k \rangle } \,.
\label{e4_17}
\ee
However, this is not so because, in particular, the corresponding counterpart of (\ref{e4_14}) becomes
a truly nonlinear system for $\{ c_1,c_2,b_1,b_2\}$ and hence cannot be easily solved. Therefore, we
continue to use the results obtained from (\ref{e4_13}). Next, a reasonable, although not the most general,
choice for $\vece_1$ is
\be
\vece_1 = \left( \ba{r} u \\ \rho_{21} v \ea \right)\,.
\label{e4_18}
\ee
Then $\vece_2$ is sought in the form
\be
\vece_2 = \left( \ba{r} \rho_{12} u \\ v \ea \right)\,,
\label{e4_19}
\ee
where $\rho_{12}$ is determined from the orthogonality condition (\ref{e4_09}):
\be
\rho_{12} = -\rho_{21} \, \frac{ \langle v,\, N_2 v \rangle }{ \langle u,\, N_1 u \rangle },
\label{e4_20}
\ee
where $N_1$ and $N_2$ are found from (\ref{e4_11}), (\ref{e4_15}), and (\ref{e4_16})
for each given value of $\rho_{21}$.

The issue is then to determine coefficient $\rho_{21}$. This can be done by imposing the requirement
that quantity (\ref{e4_17}), which is a nonlinear function of $\rho_{21}$, be maximized with respect
to that coefficient. It can be shown, with some effort, that this nonlinear optimization problem can
be solved time-efficiently, i.e. without repeated evaluation of the inner products in (\ref{e4_17}).
We performed several experiments with the Examples reported in Section 4.2 and concluded that simply taking
\be
\rho_{21}=1
\label{e4_21}
\ee
instead of solving the optimization problem for that coefficient was the optimal choice, for the
 following reasons. In many cases, we
empirically found that the ``optimal" value for $\rho_{21}$ was close to
one, and hence the considerable complexification of the code needed to compute that value did not
justify the obtained improvement of the convergence rate by just a few percent. Moreover, in some
examples we found that the iterations were initially selecting a value of $\rho_{21}$ that was not
close to (\ref{e4_21}), and then they would quickly diverge. 
(This probably occurred when the initial condition was not sufficiently close
to the exact solution.)
On the other hand, setting $\rho_{21}$
according to (\ref{e4_21}) always resulted in the convergence of the iterations. Thus we conclude
that taking the eigenvectors $\vece_{1,2}$ according to Eqs. (\ref{e4_18})--(\ref{e4_21}) constitutes
the optimal practical choice. The results presented in Section 4.2 justify the validity of this choice.

\smallskip

We now state the algorithm of the generalized \PM\ for coupled nonlinear wave equations, 
which is obtained by ``delinearizing" Eq. (\ref{e4_04}):
\be
\left( \ba{c} u \\ v \ea \right)_{n+1}  =  \left( \ba{c} u \\ v \ea \right)_n   +
 \left[ \vecN^{-1}
\left( \vecL_0 \left( \ba{c} u \\ v \ea \right)\, \right)_n -
\sum_{k=1}^2 \gamma_k
\frac{ \left\langle \vece_{k,n}, \,
              \left( \vecL_0 \left( \ba{c} u \\ v \ea \right)\, \right)_n  \right\rangle}
{ \langle \vece_{k,n}, \,  \vecN \vece_{k,n} \rangle}\, \vece_{k,n}\, \right] \D \tau \,,
\label{e4_26}
\ee
where $\vece_{k,n}$ are computed using the components $u_n,v_n$ at each iteration, and $\vecN$ and
$\gamma_k$ are computed iteratively until the solution reaches a prescribed accuracy (see Remark 3.5).
Iteration scheme (\ref{e4_26}) along with the details of calculation of $N$ and $\vece_k$ (Eqs.
(\ref{e4_11}), (\ref{e4_15}), (\ref{e4_16}), 
and (\ref{e4_18})--(\ref{e4_21})) is the main result of this
Section. As we noted in the Introduction, the reader who is not interested in implementation
issues of this algorithm may skip the remainder of this Section.

\smallskip

{\bf Remark 4.2} \ This Remark extends to the case of coupled equations the observation stated in
Remark 3.2. Namely, to calculate the coefficients of $\vecN$ and the eigenvalues $\alpha_{1,2}$ at the
$(n+1)$st iteration, one
requires the values of $\vecL\vece_{1,2}$, where $\vece_{1,2}$ are found from (\ref{e4_18}) and
(\ref{e4_19}) using the available values of $u_n$ and $v_n$. Now, the expressions
\be
\vecL\vece_k+{\rm const}\cdot\vecL_0 \left( \ba{c} u_n \\ v_n \ea \right)
\label{e4_22}
\ee
are equal to each other up to the order $O(\tu_n,\tv_n)$ for any value of the constant, and so
the issue is which of these expressions to use when computing $\vecL\vece_{1,2}$. In our simulations,
we found that computing $\vecL\vece_k$ at the $n$th iteration as
\be
\vecL\vece_{k} \, \equiv \,
(\vecL\vece_k)_n - \left( \vecL_0 \left( \ba{c} u \\ v \ea \right) \, \right)_n,
\qquad k=1,2
\label{e4_23}
\ee
(i.e. taking in (\ref{e4_22}) \ const=$-1$) results in a sufficiently broad range of initial
conditions $(u_0,v_0)$ that converge to the solitary wave $(u,v)$. For comparison, taking in (\ref{e4_22})
const=$0$ required the initial conditions to be much closer to the exact solution for the iterations
to converge. Thus, to compute $\alpha_{1,2}$ in (\ref{e4_05}),
we used the expression given by (\ref{e4_23}).
Note that while for $k=1$, this result is an obvious extension of
(\ref{e3_14}) (upon taking into account (\ref{e4_18}) and (\ref{e4_21})), for $k=2$
this result is {\em not} obvious and was arrived at upon experimentation with various
values of the constant in (\ref{e4_22}).

\smallskip

{\bf Remark 4.3} \ When system (\ref{e4_01}) is decoupled, i.e., $F_{1,v}=F_{2,u}=0$, the
approximate eigenvalues $\alpha_{1,2}$ of $\vecN^{-1}\vecL$ must be equal. Indeed,
in this case, from (\ref{e4_05}) one has:
\be
\ba{rl}
\alpha_1 = & \dst \frac{ \langle u, \, L_{11} u\rangle }{ \langle u,\, N_1 u \rangle } \cdot
 \frac{ 1 + ( \langle v, \, L_{22} v\rangle / \langle u, \, L_{11} u\rangle )}
  { 1 + ( \langle v, \, N_2 v\rangle / \langle u, \, N_1 u\rangle )} \,,
 \vspace{0.2cm} \\
\alpha_2 = & \dst \frac{ \langle u, \, L_{11} u\rangle }{ \langle u,\, N_1 u \rangle } \cdot
 \frac{ \rho_{12}^2 + ( \langle v, \, L_{22} v\rangle / \langle u, \, L_{11} u\rangle )}
  { \rho_{12}^2  + ( \langle v, \, N_2 v\rangle / \langle u, \, N_1 u\rangle )} \,.
\ea
\label{add1_4_01}
\ee
Next, using Eq. (\ref{e4_16}) with $L_{12}\equiv L_{21}=0$, one obtains:
\be
\frac{ \langle v, \, N_{2} v\rangle }{ \langle u, \, N_{1} u\rangle } \,=\,
\frac{b_2}{b_1}\, \frac{\langle v, \, (\kappa_2-\nabla^2) v\rangle }
 { \langle u, \, (\kappa_2-\nabla^2)  u\rangle }  \,=\,
 \frac{ \langle v, \, L_{22} v\rangle }{ \langle u, \, L_{11} u\rangle }\,.
\label{add1_4_02}
\ee
Substituting (\ref{add1_4_02}) into (\ref{add1_4_01}), one obtains $\alpha_1=\alpha_2$.
This fact is, thus, a consequence of the coefficients of the entries of $\vecN$
satisfying (\ref{e4_16}).

\smallskip

{\bf Remark 4.4} \ For completeness of this presentation, we note that it is possible to find such a
form of Eqs. (\ref{e4_01}) for which the coefficients of operator $\vecN$, the coefficient $\rho_{12}$,
and the eigenvalues $\alpha_{1,2}$ (and hence $\gamma_{1,2}$) can be obtained analytically (i.e.,
similarly to how the optimal $\gamma$ given by (\ref{e2_18}) is obtained for Eq. (\ref{e1_01})).
For the case of two coupled equations, we derive the corresponding class of equations in Appendix 4.
A particular equation from that class is considered in Example 4.3 below. Note that for this class
of equations, $\vecN={\rm diag}(M_{11},M_{22})$, where $M_{11}$ and $M_{22}$ 
are defined in Eq.~(\ref{e4_01}).
This is a counterpart of the relation $N=M$ for a single equation with power-law nonlinearity, 
noted after Eq.~(\ref{e3_11}).


\subsection{Examples of application of the generalized \PM\ to coupled equations}

The examples presented below are restricted to systems of two coupled stationary wave equations.
We focused on those examples where the components $u$ and $v$ of the solitary wave have distinctly
different amplitudes and widths; this is done to apply as strict as possible a test to our method.
Also, in all of these examples except Example 4.1, the computational domain was a square with the
side of $8\pi$ and $2^7$ mesh points along each side. In Example 4.1, the side of the square was
$12\pi$ with $2^8$ mesh points per side.

\smallskip

{\bf Example 4.1} \ Consider a vector generalization of the equation from Example 3.1:
\be
\ba{l}
\nabla^2 u + 4(\cos^2x + \cos^2y )u + u (u^2+\sigma v^2 ) \,=\, \mu_1 u \vspace{0.1cm} \\
\nabla^2 v + 4(\cos^2x + \cos^2y )v + v (\sigma u^2+ 4 v^2 ) \,=\, \mu_2 v.
\ea
\label{e4_27}
\ee
Here the asymmetry between $u$ and $v$ is provided by two sources: \ (i) by the different coefficients,
`$1$' and `$4$', in front of the self-nonlinearity terms and, more importantly, \ (ii) by the different
propagation constants $\mu_1$ and $\mu_2$. Specifically, we used
$$
\mu_1=4.95 \quad {\rm and} \quad \mu_2=6.5.
$$
In the absence of coupling ($\sigma=0$), this corresponds to the solution $u$ being near the edge of the
zeroth band gap and $v$ being suficiently far away from that edge. Consequently, $v$ is significantly
``taller" and more localized than $u$ (see, e.g., \cite{EfremidisH03}). When the coupling is present
($\sigma>0$), the structure of the composite solution remains qualitatively the same;
such a solution for
\be
\sigma=0.5
\label{e4_28}
\ee
is plotted in Fig.~\ref{fig4}. Starting with the initial condition
\be
\ba{l}
\dst   u_0=A_1 \, e^{-(x^2+y^2)/W_1^2} \vspace{0.1cm} \\
\dst   v_0=A_2 \, e^{-(x^2+y^2)/W_2^2},
\ea
\label{e4_29}
\ee
where $A_{1,2}$ and $W_{1,2}$ are listed in Table 1, the iterations (\ref{e4_26}) with $\D \tau=1$
\footnote{This value of $\D \tau$ is likely not to be optimal (see Example 3.1). However, our focus here
is not to optimize the convergence rate but to demonstrate the validity of the method.}
takes about 710 iterations to converge to accuracy of $10^{-10}$. Here and below, the accuracy for
two-component solitary waves is defined similarly to (\ref{e1_12}):
\be
E_n= \left( \frac{ \langle u_n-u_{n-1}, u_n-u_{n-1} \rangle}{ \langle u_n, u_n \rangle }
 + \frac{ \langle v_n-v_{n-1}, v_n-v_{n-1} \rangle}{ \langle v_n, v_n \rangle }
  \right)^{1/2}\,.
\label{e4_30}
\ee

In all of the examples of this Section, we monitored the following quantities: coefficient $\rho_{12}$
(see (\ref{e4_19})--(\ref{e4_21})); factors
\be
I_k = \frac{ \langle \vecN\vece_k, \, \vecL\vece_k \rangle^2 }
 { \langle \vecN\vece_k, \, \vecN\vece_k \rangle \langle \vecL\vece_k, \, \vecL\vece_k \rangle },
\qquad k=1,2,
\label{e4_31}
\ee
which show how close vectors $\vece_k$ are to the true eigenvectors of \ $\vecN^{-1}\vecL$; the
eigenvalues $\alpha_{1,2}$ (see (\ref{e4_06})); and the coefficients $c_1,c_2,b_2$ of $\vecN$
(we set $b_1=1$ without loss of generality). These quantities are reported in Table 1.
In particular, one sees that $\vece_1$ is a closer approximation to its corresponding true eigenvector
of $\vecN^{-1}\vecL$ than $\vece_2$ is to its true eigenvector; this is expected since the coefficients
of operator $\vecN$ are computed using $\vece_1$. Note, however, from the reported values of $I_{1,2}$,
that both $\vece_1$ and $\vece_2$ approximate their respective true eigenvectors quite well.

To benchmark the performance of the method, we also obtained the solution of the uncoupled system,
(\ref{e4_27}) with $\sigma=0$, in two ways. First, we used the vector generalization of the \PM\
described in Section 4.1. For $\D \tau=1$, the iterations converged to accuracy $10^{-10}$ in about
950 iterations.
(Let us note, in passing, that the numerically found $\alpha_{1,2}$ agree with Remark 4.6.)
As the second way of obtaining the same solutions, we solved each
of the uncoupled equations (\ref{e4_27}) using the generalized \PM\ for a single equation, as described
in Section 3.1. The iterations for components $u$ and $v$ took, respectively, about 950 and 80
iterations to converge to the accuracy of $10^{-10}$. Comparing this with the number of iterations
needed to obtain the solution of the uncoupled system via the first method, we conclude that the
convergence rate of the vector form of the generalized \PM\ is determined by such a rate for the more
slowly converging component of the solitary wave.

\smallskip

{\bf Example 4.2} \ We now consider a system of linearly coupled nonlinear Schr\"odinger equations:
\be
\ba{l}
\nabla^2 u + u^3 + \sigma v = u  \vspace{0.1cm} \\
\nabla^2 v + v^3 + \sigma u = v\,.
\ea
\label{e4_33}
\ee
This system extends to two dimensions the equations of the so-called
nonlinear directional coupler \cite{AkhmedievA93}. In one spatial
dimension, these equations are known to possess symmetric ($u=v$),
anti-symmetric ($u=-v$), and, for $\sigma<0.6$, asymmetric ($|u|\neq
|v|$) solitary waves \cite{AkhmedievA93}. The smaller the $\sigma$,
the greater the asymmetry between the two components of the latter
solution. To our knowledge, asymmetric solutions of the
two-dimensional system (\ref{e4_33}) have not been reported
previously.

We considered Eqs. (\ref{e4_33}) with $\sigma=0.5$. By trial and error, we found that the initial condition
(\ref{e4_29}) with the parameters reported in Table 1
leads the iterations to converge to the solution depicted in Fig.~\ref{fig6}.
(It should be noted that this initial condition must be quite close to the exact
solution in order for the iterations to converge. For example, if one takes $A_2=0.4$ or $A_2=0.6$
instead of $0.5$, as in Table 1, then the iterations converge to either the symmetric or anti-symmetric
solitary wave.)
Next, by running the simulations and monitoring,
at each iteration, the approximate eigenvalues $\alpha_{1,2}$, we observed that $\alpha_2$ is a large
negative number (see Table 1). Then, to satisfy the necessary convergence condition (\ref{e2_24}),
one needs to use a rather small step size $\D \tau$. By trial and error, we found that $\D \tau=0.08$ results
in nearly the fastest convergence of the method for system (\ref{e4_33}).

Note that since $\alpha_2<0$ in this example, the iterations would still converge if $\gamma_2$ 
were set to zero.

\smallskip

{\bf Example 4.3} \ As the last example, we applied method (\ref{e4_25}) to a system of equations
that describe copropagation of the fundamental and second harmonic fields in an optical medium with
quadratic nonlinearity:
\be
\ba{l}
\dst   \nabla^2 u + uv = \mu_1 u \vspace{0.1cm} \\
\dst   \nabla^2_{\delta} v + \frac12 u^2 = \mu_2 v,  \qquad
\nabla^2_{\delta} \equiv \partial^2_x + \delta \partial^2_y\,.
\ea
\label{e4_34}
\ee
Multidimensional solutions of this and
related systems were considered in quite a few studies; see, e.g., a recent paper \cite{LloydC04}
and references therein.
It should be noted that Eqs. (\ref{e4_34}) are a special case of the
system of two coupled wave equations for which all the coefficients in the \PM\ can be determined
analytically (see Eq. (\ref{A2_15}) in Appendix 4). In particular, as follows from the last paragraph
of Appendix 4, one should have $c_{1,2}=\mu_{1,2}$ and $b_2=1$.
Thus, this example provides a test of whether our method would obtain
these coefficients correctly, and it indeed did so. Specifically, we took
\be
\delta=10, \quad \mu_1=1.5, \quad \mu_2=9;
\label{e4_35}
\ee
then one can see that the values of $c_{1,2}$ and $b_2$,
reported in Table 1, are indeed as stated above.
Moreover, the values of $\rho_{12}$ and $\alpha_{1,2}$ agree with those given by Eq.
(\ref{A2_13}) in Appendix 4. Finally, we note that the formulae for the
calculation of $c_{1,2}$, $b_2$, $\rho_{12}$, and $\alpha_{1,2}$ are
those given in Section 4.1 with one modification: all occurrences of
$\nabla^2 v$ should be replaced with $\nabla^2_{\delta}v$. The
corresponding solitary wave is shown in Fig.~\ref{fig7}.


\section{Summary}

In this work, we obtained the following two main results.

First, in Section 3, we extended the well-known Petviashvili iteration method to find solitary wave solutions
of a broad class of Hamiltonian nonlinear wave equations with {\em arbitrary} form of nonlinearity and
potential function; see Eq. (\ref{e3_01}). 
Our algorithm is given by Eqs. (\ref{e3_06})--(\ref{e3_08}), (\ref{e3_11}), and (\ref{e3_12}).
The generalized \PM\ can be applied even when the equation is
nonlocal; see Example 3.3. The computational cost of this method only slightly exceeds that of the original
\PM, since the (few) parameters required to carry out the iterations need to be computed only until the
solution reaches some relatively low accuracy; see Remark 3.5.

Second, in Section 4, we extended this method to systems of coupled Hamiltonian wave equations. Our main
result here was the finding of a way in which all the required parameters of the iteration scheme can be
computed by explicit expressions, obtained from solving a simple {\em linear} system of algebraic equations.
The algorithm (for two equations) is given by Eqs. (\ref{e4_26}), 
(\ref{e4_11}), (\ref{e4_15}), (\ref{e4_16}), 
and (\ref{e4_18})--(\ref{e4_21})). 

Appendices 1 and 2 contain, respectively, a Matlab code illustrating the algorithm of Section 3 and
a pseudocode for the algorithm of Section 4. Appendix 3 contains an extension of the
algorithm of Section 4 to three (and more) equations. Finally, Appendix 4 contains
a collateral result:
the form of a system of two coupled equations for which the parameters of our generalized
Petviashvili iteration scheme can be found analytically (as in the original \PM\ for a single equation
with power-law nonlinearity).

\section*{Acknowledgement}
The work of T.I.L was supported in part by the National Science Foundation under
grant DMS-0507429, and the work of J.Y. was supported in part by The Air Force Office of
Scientific Research under grant USAF 9550-05-1-0379.


\setcounter{equation}{0}
\renewcommand{\theequation}{A1.\arabic{equation}}
\section*{Appendix 1: \ Matlab code for Example 3.1}

\begin{verbatim}
N=2^7;  d=10*pi/N;                    % mesh sizes along x and y
x=[-5*pi:d:5*pi-d]; y=x;
[X,Y]=meshgrid(x,y);                  % 2D x- and y-arrays
kx=2*pi/(10*pi)*[0:N/2-1  -N/2:-1];  ky=kx;
[KX,KY]=meshgrid(kx,ky);   K2=KX.^2+KY.^2;
Dt=0.4;                               % Delta tau
mu=3.7;                               % prop. constant of the soliton
W=3*((cos(X)).^2+(cos(Y)).^2)-mu;     % V(x)-mu
u0=1.5*exp(-(X.^2+Y.^2));  u=u0;      % initial condition
norm_Du=1;                            % initialize E_n defined in (1.11)
while norm_Du >= 10^(-10)
    u_old=u;   fftu=fft2(u);   ucube=u.^3;   DEL_u=real(-ifft2(K2.*fftu));
    if norm_Du >= 10^(-3)  % when E_n > 10^(-3), compute c and gamma 
        dVu=2*ucube;   u_u=sum(sum(u.^2)); 
        u_Lu=sum(sum(dVu.*u));   DELu_DELu=sum(sum(DEL_u.^2)); 
        DELu_Lu=sum(sum(dVu.*DEL_u));   u_DELu=sum(sum(u.*DEL_u));
        c=(u_Lu*DELu_DEL_u-DELu_Lu*u_DELu)/(u_Lu*u_DELu-DELu_Lu*u_u)
        u_Nu=c*u_u-u_DELu;   alpha=u_Lu/u_Nu;   gamma=1+1/(alpha*Dt);
        fftNinv=1./(c+K2);             % Fourier symbol of N^(-1)
    else       % once E_n < 10^(-3), use previously computed c and gamma.
        u_Nu=sum(sum(c*u.^2-u.*nabla2_u));
    end      
    L0u=DEL_u + W.*u + ucube;
    u=u+Dt*real(ifft2(fft2(L0u).*fftNinv)-u*gamma*sum(sum(u.*L0u))/u_Nu);
    norm_Du=sqrt(sum(sum((u-u_old).^2))*d^2);   % new E_n
end
\end{verbatim}


\setcounter{equation}{0}
\renewcommand{\theequation}{A2.\arabic{equation}}
\section*{Appendix 2: \ Pseudocode for time-efficient implementation of algorithm (\ref{e4_26})}

Here we suggest an order in which various quantities, required to perform each iteration
in algorithm (\ref{e4_26}), can be computed. Computing these quantities
in this order allows one to avoid repeated time-intensive evaluations, e.g., of inner products
such as those required in (\ref{e4_05}), an so on.

For notational convenience, we denote, in this Appendix only,
\be
u_1\equiv u_n, \qquad u_2 \equiv v_n,
\label{A1_01}
\ee
where $(u_n,v_n)$ are the solution's components at the $n$th iteration. This notation will facilitate
the extension of the steps listed below to the case of more than two coupled equations.
(A minor modification of this algorithm occurring for more than two equations is
described in Remark 4.5.)
In the notations used below, any index (e.g., $j$) is assumed to take on the values
from one to the number of equations (two in the case considered in this paper). The summation
indices (e.g., $k$ in $\sum_k$) run over the same range of values.

The first column of the list(s) below shows which quantity is computed at the given step.
The second column shows, which equations of the main text and results of which previous steps of this list,
are used at the given step.

\bigskip

{\em The first block of step, listed below,
is performed at each iteration, irrespective of the magnitude of the error.}
\setlength\extrarowheight{6pt}
\bea
\hspace*{-4cm} &  \nabla^2 u_k  &  \qquad \{ \ref{A1_01} \}
 \label{A1_02} \\
\hspace*{-4cm}  & (\vecL_0)_{jk}u_k   & \qquad \{ \ref{A1_02} \}
 \label{A1_03} \\
\hspace*{-4cm}  & \sum_k (\vecL_0)_{jk}u_k   & \qquad  \{ \ref{A1_03} \}
 \label{A1_04} \\
\hspace*{-4cm}  & \langle u_j, \, \sum_k (\vecL_0)_{jk}u_k \rangle  & \qquad  \{ \ref{A1_04} \}
 \label{A1_05}
\eea

\bigskip

{\em
The second block of steps, listed below,
contains steps that are required for the calculation of the parameters of operator
$\vecN$, the eigenvectors $\vece_k$, and the parameters $\gamma_k$. These steps need to be performed
only while the error is greater than a user-defined threshhold (e.g., $10^{-3}$); see Remark 3.6 and
a note after Eq. (\ref{e4_16}).
}
\setlength\extrarowheight{6pt}
\bea
\hspace*{-4cm} &  L_{jk} u_k  &   \qquad  \{ \ref{e4_12}, \; \ref{A1_02}  \}
 \label{A1_06} \\
\hspace*{-4cm}  & \langle u_j, \, \sum_k L_{jk}u_k \rangle  & \qquad  \{ \ref{A1_06} \}
 \label{A1_07} \\
\hspace*{-4cm}  & \langle \nabla^2 u_j, \, \sum_k L_{jk}u_k \rangle  &
  \qquad  \{ \ref{A1_02}, \; \ref{A1_06} \}
 \label{A1_08} \\
\hspace*{-3cm}  & \langle u_k, \, u_k \rangle, \;\; \langle u_k, \, \nabla^2 u_k \rangle, \;\;
 \langle \nabla^2 u_k, \, \nabla^2 u_k \rangle  & \qquad  \{ \ref{A1_02}  \}
 \label{A1_09} \\
\hspace*{-4cm}  & \langle \nabla^2 u_j, \, \sum_k (\vecL_0)_{jk}u_k \rangle  &
 \qquad  \{ \ref{A1_02} , \; \ref{A1_04} \}
 \label{A1_10} \\
\hspace*{-4cm}  & \langle \sum_m (\vecL_0)_{jm}u_m , \, \sum_k (\vecL_0)_{jk}u_k \rangle  &
 \qquad  \{ \ref{A1_04} \}
 \label{A1_11} \\
\hspace*{-4cm}  & \langle L_{jm}u_m , \, L_{jk}u_k \rangle  &
 \qquad  \{ \ref{A1_06}  \}
 \label{A1_12} \\
\hspace*{-4cm}  & \langle L_{jm}u_m , \, \sum_k (\vecL_0)_{jk}u_k \rangle  &
 \qquad  \{ \ref{A1_04} , \; \ref{A1_06} \}
 \label{A1_13} \\
\hspace*{-4cm}  & \langle u_j , \, \sum_k L_{jk}e_{k1} \rangle  &
 \qquad  \{ \ref{e4_12}, \; \ref{e4_23}, \; \ref{A1_05} , \; \ref{A1_07} \}
 \label{A1_14} \\
\hspace*{-4cm}  & \langle \nabla^2 u_j , \, \sum_k L_{jk}e_{k1} \rangle  &
 \qquad  \{ \ref{e4_12}, \; \ref{e4_23}, \; \ref{A1_08} , \; \ref{A1_10} \}
 \label{A1_15} \\
\hspace*{-4cm}  & \kappa_k &
 \qquad  \{ \ref{e4_15}, \; \ref{A1_09}, \; \ref{A1_14} , \; \ref{A1_15} \}
 \label{A1_16} \\
\hspace*{-4cm}  & b_k, \; c_k \;\; \mbox{(take $b_1=1$)} &
 \qquad \{ \ref{e4_16}, \; \ref{e4_15}, \; \ref{A1_09}, \; \ref{A1_14} , \; \ref{A1_16} \}
 \label{A1_17} \\
\hspace*{-4cm}  & \langle u_k , \, N_k u_k  \rangle  &
 \qquad  \{ \ref{e4_11}, \; \ref{A1_09}, \; \ref{A1_16} , \; \ref{A1_17} \}
 \label{A1_18} \\
\hspace*{-4cm}  & \rho_{12}   &
 \qquad  \{ \ref{e4_20}, \; \ref{A1_18} \}
 \label{A1_19} \\
\hspace*{-4cm}  & \langle \vecN \vece_k , \, \vecN \vece_k \rangle  &
 \qquad  \{ \ref{e4_18}, \; \ref{e4_19}, \; \ref{e4_21}, \; \ref{A1_18}, \; \ref{A1_19} \}
 \label{A1_20} \\
\hspace*{-4cm}  & \langle \vecL \vece_k , \, \vecL \vece_k \rangle  &
 \qquad  \{ \ref{e4_23}, \; \ref{A1_12}, \; \ref{A1_13}, \; \ref{A1_19} \}
 \label{A1_21} \\
\hspace*{-4cm}  & \langle u_j , \, \sum_k L_{jk} e_{k2}  \rangle  &
 \qquad  \{ \ref{e4_19}, \; \ref{e4_23}, \; \ref{A1_05}, \; \ref{A1_07} , \; \ref{A1_19} \}
 \label{A1_22} \\
\hspace*{-4cm}  & \langle \nabla^2 u_j , \, \sum_k L_{jk} e_{k2}  \rangle  &
 \qquad  \{ \ref{e4_19}, \; \ref{e4_23}, \; \ref{A1_08}, \; \ref{A1_10} , \; \ref{A1_19} \}
 \label{A1_23} \\
\hspace*{-4cm}  & \langle \vecN \vece_k , \, \vecL \vece_k \rangle  &
 \qquad  \{ \ref{A1_14}, \, \ref{A1_15}, \,\ref{A1_17}, \,\ref{A1_19}, \,\ref{A1_22}, \,\ref{A1_23} \}
 \label{A1_24} \\
\hspace*{-4cm}  & \langle \vece_k , \, \vecN \vece_k \rangle  &
 \qquad  \{ \ref{A1_18}, \; \ref{A1_19} \}
 \label{A1_25} \\
\hspace*{-4cm}  & \langle \vece_k , \, \vecL \vece_k \rangle  &
 \qquad  \{ \ref{A1_14}, \; \ref{A1_19}, \; \ref{A1_22} \}
 \label{A1_26} \\
\hspace*{-4cm}  & \alpha_k, \;\; \gamma_k  &
 \qquad  \{ \ref{e4_05}, \; {\rm possibly} \; \ref{e3_17}, \; \ref{A1_25}, \; \ref{A1_26} \}
 \label{A1_27}
\eea

\bigskip

{\em
The last block of steps
is again performed at each iteration, irrespective of the magnitude of the error.
Note that the latest computed results from the second block are used in this one,
whereever they are required.}
\setlength\extrarowheight{6pt}
\bea
\hspace*{-4cm} &  \langle  \vece_j , \, \sum_k (\vecL_0)_{jk} u_k  \rangle  &
\qquad \{ \ref{A1_05}, \; \ref{A1_19} \}
 \label{A1_28} \\
\hspace*{-3cm} &  u_k \; \mbox{at next iteration} &
\qquad \{ \ref{e4_26}, \; \ref{A1_04}, \; \ref{A1_17}, \ref{A1_19}, \; \ref{A1_25}, \; \ref{A1_27}, \;
 \ref{A1_28}  \}
 \label{A1_29}
\eea


\setcounter{equation}{0}
\renewcommand{\theequation}{A3.\arabic{equation}}
\section*{Appendix 3: \ Extension of the algorithm of Section 4.1 to any number of coupled equations}

For simplicity, we present the details for the case of three equations; for 
more equations, this treatment can be extended straightforwardly.
The counterparts of Eqs. (\ref{e4_09})
and (\ref{e4_11}) for three equations are, respectively:
$$
\langle \vece_j, \, \vecN\vece_k \rangle =0, \qquad j,k=1,2,3, \quad j\neq k
\eqno (\ref{e4_09}')
$$
and
$$
\vecN = {\rm diag}(N_1,N_2,N_3), \qquad N_k=c_k-b_k\nabla^2, \quad k=1,2,3.
\eqno (\ref{e4_11}')
$$
Then, Eqs. (\ref{e4_15}) with $k=1,2,3$ are unchanged, and Eq. (\ref{e4_16}) is replaced with
analogous expressions for $b_k/b_1$ where index ``2" in (\ref{e4_16}) is replaced with $k=2,3$.
Finally, Eqs. (\ref{e4_18}) and (\ref{e4_19}) are replaced by
$$
\vece_1= \left( \ba{c} u \\ v \\ w \ea \right), \quad
\vece_2= \left( \ba{r} \rho_{12} u \\ v \\ \rho_{32} w \ea \right), \quad
\vece_3= \left( \ba{r} \rho_{13} u \\ \rho_{23} v \\  w \ea \right),
\eqno (\ref{e4_18}') 
$$
where $w$ is the third component of the solitary wave. Since the orthogonality conditions
(\ref{e4_09}$'$) yield only three constraints for the four coefficients $\rho_{jk}$, we impose an
additional arbitrary constraint, which we take to be simply
\be
\rho_{32}=0.
\label{e4_24}
\ee
Then Eqs. (\ref{e4_09}$'$), (\ref{e4_11}$'$), and (\ref{e4_24}) yield Eq. (\ref{e4_20}) 
(with $\rho_{21}=1$)
for $\rho_{12}$ and the following system for $\rho_{13}$ and $\rho_{23}$:
\be
\ba{l}
\rho_{13} \langle u, N_1 u \rangle + \rho_{23} \langle v, N_2 v \rangle \,=\,
 - \langle w, \, N_3 w \rangle   \vspace{0.1cm} \\
\rho_{13}\rho_{12} \langle u, N_1 u \rangle + \rho_{23} \langle v, N_2 v \rangle \,=\, 0\,.
\ea
\label{e4_25}
\ee
System (\ref{e4_25}) can always be solved because $\rho_{12}\neq 1$.


\setcounter{equation}{0}
\renewcommand{\theequation}{A4.\arabic{equation}}
\section*{Appendix 4: \ Extension of the original \PM\ to two coupled equations}

Here we will derive the form of two coupled equations for which there exist explicit analytical
expressions for the coefficients $\alpha_{1,2}$ etc. (see Remark 4.7).

Using the analogy with the case of a single equation for which the constant $\gamma$ in the
original \PM\ is given by the explicit formula (\ref{e2_18}), we seek the two coupled equations in
question in the form:
\be
 \vecL_0 \left( \ba{c} u \\ v \ea \right) \; \equiv \;
 - \left(  \ba{cc} M_{11} & 0 \\ 0 & M_{22} \ea \right)
 \left( \ba{c} u \\ v \ea \right) +
 \left( \ba{c} \sum_{j} a_{1j} \left( u^{p_{1j}} v^{q_{1j}} \right) \\
 \sum_{j} a_{2j} \left( u^{p_{2j}} v^{q_{2j}} \right) \ea    \right)
 \; = \; \left( \ba{c} 0 \\ 0 \ea \right)\,,
\label{A2_01}
\ee
where $M_{11}$ and $M_{22}$ are self-adjoint positive
definite operators, as before; $p_{kj}$ and $q_{kj}$, $k=1,2$, are
some constants; and $a_{kj}$ are linear operators (in particular,
they may be constants). As we pointed out after Eq.~(\ref{e4_01}), a more
general Hamiltonian system with off-diagonal terms $M_{12}v$ and
$M_{12}u$ in the matrix above, can be reduced to form (\ref{A2_01}) by
a linear transformation of $u$ and $v$. 
The {\em key condition} which will allow us
to determine the relation between the exponents $p_{kj}$ and
$q_{kj}$ as well as the parameters of the \PM, is that there is {\em
no algebraic relation} (such as, e.g., $u=const\cdot v$) between the components $u$ and $v$ of the
solitary wave.

First, we require that the linearized operator $\vecL$
of this equation be self-adjoint (see (\ref{e4_03})),
which yields that for {\em each} $j$ (see the key condition above), either
\bea
q_{1j} a_{1j} & = & p_{2j}a_{2j} ,
\label{A2_02} \\
p_{1j} & = & p_{2j}-1,
\label{A2_03} \\
q_{1j} & = & q_{2j}+1,
\label{A2_04}
\eea
or
\be
q_{1j}=0 \qquad {\rm and} \qquad p_{2j}=0.
\label{A2_05}
\ee
Next, we require that equation
\be
\vecL \left( \ba{c} u \\ v \ea \right) = \alpha_1 \vecN \left( \ba{c} u \\ v \ea \right)
\label{A2_06}
\ee
be satisfied. In view of the key condition, and since (\ref{A2_06}) is to be satisfied exactly,
it is intuitively clear (and can be easily shown) that the only possibility for
operator $\vecN$ is:
\be
M_{12}=0, \quad \so \quad \vecN =
\left(  \ba{cc} M_{11} & 0 \\ 0 & b M_{22} \ea \right) ,
\label{A2_07}
\ee
where for the moment constant $b$ is arbitrary.
%
%
Then (\ref{A2_06}) and the key condition yield
\be
\ba{rcl}
p_{1j}+q_{1j}-1 & = & \alpha_1 \\
p_{2j}+q_{2j}-1 & = & \alpha_1 b
\ea   \qquad \mbox{for all $j$}.
\label{A2_09}
\ee
The counterpart of (\ref{A2_06}) for the eigenvector $\vece_{2}$ (see (\ref{e4_06}) and (\ref{e4_19}))
yields a similar system:
\be
\ba{rcl}
\rho_{12}(p_{1j}-1)+q_{1j} & = & \alpha_2  \rho_{12} \\
\rho_{12} p_{2j}+q_{2j}-1 & = & \alpha_2 b
\ea   \qquad \mbox{for all $j$}.
\label{A2_10}
\ee
Eliminating $p_{2j}$, $q_{2j}$, and $\alpha_1$ from (\ref{A2_03}), (\ref{A2_04}), and (\ref{A2_09})
shows that the following two subcases are possible:
\be
\ba{rcl}
(a): & \qquad  & p_{1j}+q_{1j} \neq 1 \quad \mbox{for all $j$, \ and $b=1$}; \\
(b): & \qquad  & p_{1j}+q_{1j}=1 \quad \mbox{for all $j$};
\ea
\label{A2_11}
\ee
note that in subcase (b), coefficient $b$ is undetermined. Proceeding with subcase (a),
we substitute Eqs. (\ref{A2_03}) and (\ref{A2_04}) into (\ref{A2_10}) and obtain:
\be
\ba{rcl}
\rho_{12}(p_{1j}-1) + q_{1j} & = & \alpha_2  \rho_{12} \\
\rho_{12}(p_{1j}+1) + q_{1j}-2 & = & \alpha_2 .
\ea
\label{A2_12}
\ee
Since this system is to hold for all $j$, with $\rho_{12}$ and $\alpha_2$ being independent of $j$,
one concludes that this system can be satisfied only for one set of values $\{ p_{1j}, q_{1j} \}$,
which we therefore redenote as $\{p,q\}$. Solving then Eqs. (\ref{A2_12}) for $\rho_{12}$ and $\alpha_2$
and Eqs. (\ref{A2_09})  for $\alpha_1$ yields:
\be
\rho_{12}\,=\,-\frac{q}{p+1}, \qquad \alpha_1=p+q-1, \qquad \alpha_2=-2.
\label{A2_13}
\ee
Also, from (\ref{A2_02})--(\ref{A2_04}) and (\ref{A2_13}) one has:
\be
a_2=-\rho_{12}\,a_1\,.
\label{A2_14}
\ee
Now, using the last equation, one verifies that the orthogonality condition (\ref{e4_09}) is satisfied
in this case. Thus, the system of two coupled equations for which the parameters $\rho_{12}$ and
$\alpha_{1,2}$ can be determined explicitly (by Eqs. (\ref{A2_13})) is given by:
\be
 - \left(  \ba{cc} M_{11} & 0 \\ 0 & M_{22} \ea \right)
 \left( \ba{c} u \\ v \ea \right) +
 \left( \ba{c} a \left( u^{p} v^{q} \right) \\
 (q/(p+1))\, a \left( u^{p+1} v^{q-1} \right) \ea    \right)
 \; = \; \left( \ba{c} 0 \\ 0 \ea \right)\,,
\label{A2_15}
\ee
where $a$ is any linear operator and $p,q$ are constants.

Similarly, one can show that subcase (b) of (\ref{A2_11}) yields the same equation (\ref{A2_15}), where
$q=1-p$. Setting the value of the free coefficient $b$ to one yields relations (\ref{A2_13}) and
(\ref{A2_14}) in this subcase as well.

Finally, one can straightforwardly verify that the case given by Eqs. (\ref{A2_05}) corresponds to
two uncoupled equations of the form (\ref{e1_01}). Thus, the only nontrivial case
in which the parameters of the \PM\ for a system of two coupled equations can be determined explicitly
is given by Eq. (\ref{A2_15}).
(As we stated after Eq. (\ref{A2_01}), it is assumed that there is no algebraic
relation between the components of the soltary wave.)
The parameters of the method are given by Eqs. (\ref{A2_13}), and
$\vecN$ is given by (\ref{A2_07}) with $b=1$; that is, $\vecN$ coincides with the linear operator
in (\ref{A2_15}), similarly to what occurs in the case of a single equation with power-law nonlinearity,
originally considered by Petviashvili \cite{Petviashvili76}.
Note that for Eq.~(\ref{A2_15}), it is not actually necessary to use the eigenvector
$\vece_2$ in algorithm (\ref{e4_26}) (i.e., one can set $\gamma_2=0$), because
$\alpha_2<0$ and the corresponding component of the error would decay on its own
(provided that the step size $\D \tau$ satisfies the constraint (\ref{e2_24})).

\vskip 2 cm

{\bf Table 1} \ \
Values of the parameters, noted around Eqs. (\ref{e4_29}) and (\ref{e4_31}) in the text,
for Examples 4.1--4.3. The asterisk next to the value of $\D \tau$ means that this time step is
close to optimal. The numbers of iterations are rounded to the nearest ten.

\vskip 1 cm

\setlength\extrarowheight{4pt}
\hspace*{-1.3cm}
\begin{tabular}{|c|c|c|c|c|c|c|c|c|}
\hline
Equation & $I_{1,2}$ & $\rho_{12}$ & $\alpha_{1,2}$ &
 $\hspace*{-0.2cm} \ba{c} c_{1,2} \\ b_2 \ea \hspace*{-0.2cm} $ & $A_{1,2}$ & $W_{1,2}^2$ &
 $ \D \tau$ &  $ \hspace*{-0.3cm} \ba{c} {\rm Number \; of} \\ {\rm iterations} \ea  \hspace*{-0.3cm} $
 \\ \hline \hline
$\ba{c} (\ref{e4_27}), \\ \sigma=0.5 \ea$
  & $0.99,\;0.69$ & $-66.2$ & $0.136, \; 0.0231$ &
$\hspace*{-0.2cm} \ba{c} 1.03,\; 14.9 \\ 7.57 \ea \hspace*{-0.2cm} $ &
 0.6, 1.5 & 2.0, 0.4 & 1.0 & 710  \\ \hline
$\ba{c} (\ref{e4_27}), \\ \sigma=0 \ea$
  & $0.98,\;0.78$ & $-12.4$ & $0.0943, \; 0.0943$ &
$\hspace*{-0.2cm} \ba{c} 1.52,\; 21.5 \\ 11.0 \ea \hspace*{-0.2cm} $ &
 0.8, 1.5 & 1.0, 0.4 & 1.0 & 950  \\ \hline
$\ba{c} (\ref{e4_33}), \\ \sigma=0.5 \ea$
&  $1.00,\;0.74$ & $-1.06\cdot 10^{-2}$ & $2.08, \; -10.1$ &
 $\hspace*{-0.2cm} \ba{c} 0.750,\; 0.500 \\ 0.162 \ea \hspace*{-0.2cm} $ &
 2.0, 0.5 & 0.7, 0.3 & $0.08^*$ & 580  \\ \hline
$\ba{c} (\ref{e4_34}), \\ (\ref{e4_35}) \ea$
 & $1.00,\;1.00$ & $-0.500$ & $1.00, \; -2.00$ &
  $\hspace*{-0.2cm} \ba{c} 1.50,\; 9.00 \\ 1.00 \ea \hspace*{-0.2cm} $ &
 1.0, 1.0 & 2.0, 2.0 & $0.7^*$ & 90  \\ \hline
\end{tabular}

\newpage

\begin{figure}[h]
\vspace*{-5cm}
\rotatebox{0}{\resizebox{10cm}{13cm}{\includegraphics[0in,0.5in]
 [8in,10.5in]{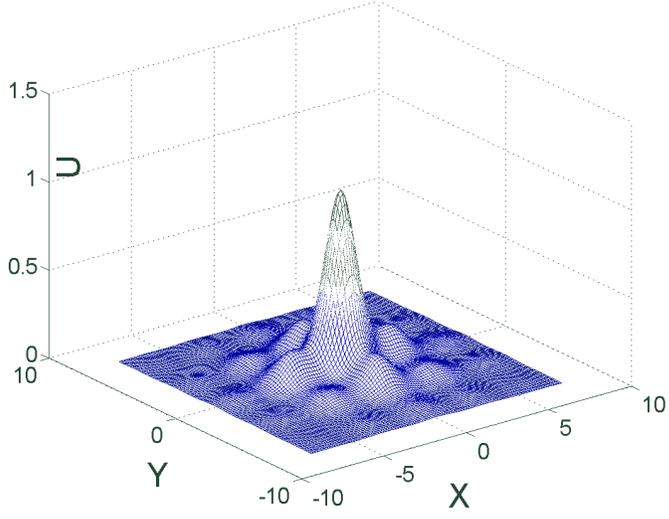}}}
\vspace{1cm}
\caption{Solution of Eq. (\ref{e3_21}) with $V_0=3$ and $\mu=3.7$.
}
\label{fig1}
\end{figure}

\vskip 2 cm

\begin{figure}[h]
\vspace*{-4cm}
\rotatebox{0}{\resizebox{10cm}{13cm}{\includegraphics[0in,0.5in]
 [8in,10.5in]{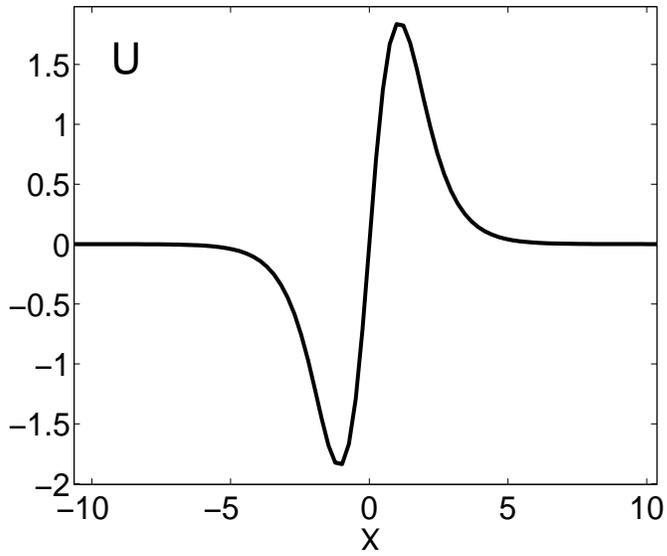}}}
\vspace{-2cm}
\caption{ Anti-symmetric solutions of Eq. (\ref{add3_01}) with $\mu=1.43$.
}
\label{fig2}
\end{figure}

\newpage

\vspace*{-1cm}
\begin{figure}[h]
\rotatebox{0}{\resizebox{10cm}{13cm}{\includegraphics[0in,0.5in]
 [8in,10.5in]{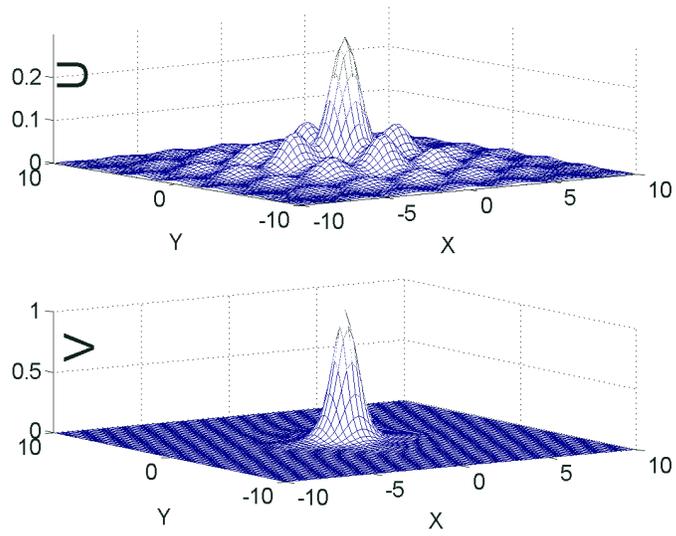}}}
\vspace{1cm}
\caption{Solution of Eqs. (\ref{e4_27}) with $\mu_1=4.95$, $\mu_2=6.5$, and $\sigma=0.5$.
Note the different vertical scales of the $u$- and $v$-components.
}
\label{fig4}
\end{figure}

\vskip 2 cm

\begin{figure}[h]
\rotatebox{0}{\resizebox{10cm}{13cm}{\includegraphics[0in,0.5in]
 [8in,10.5in]{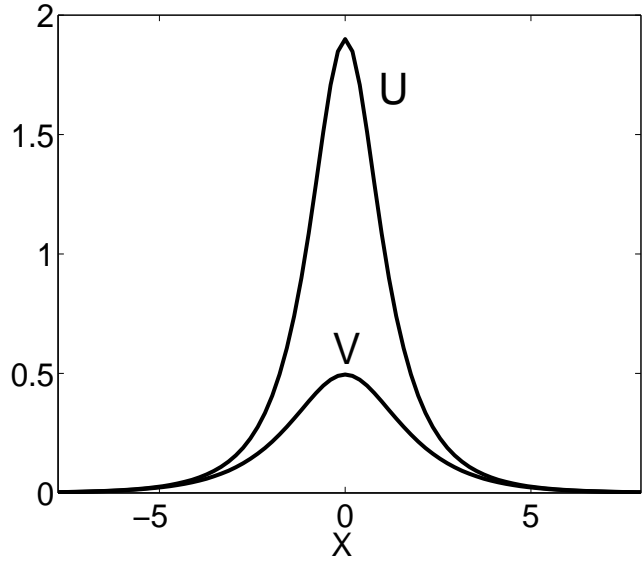}}}
\vspace{-2cm}
\caption{Solution of Eqs. (\ref{e4_33}) with
$\sigma=0.5$ along the $x$-axis (the solution is radially symmetric).
}
\label{fig6}
\end{figure}

\newpage

\begin{figure}[h]
\vspace*{-4cm}
\rotatebox{0}{\resizebox{10cm}{13cm}{\includegraphics[0in,0.5in]
 [8in,10.5in]{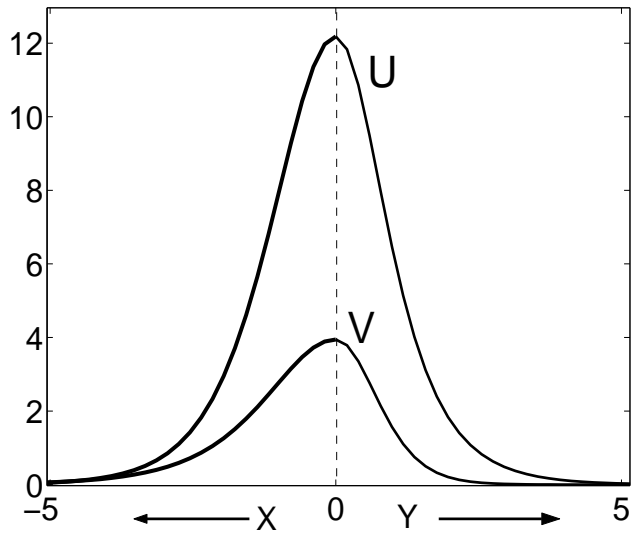}}}
\vspace{-2cm}
\caption{Solution of Eqs. (\ref{e4_34}), (\ref{e4_35}) along the $x$-axis (to the left of the
dashed line) and the $y$-axis (to the right of the dashed line).
}
\label{fig7}
\end{figure}

\end{document}